\documentclass[english,prb,aps,showpacs,twocolumn]{revtex4-1}

\usepackage{amsmath,amssymb,mathrsfs}
\usepackage{graphicx}
\usepackage{xcolor}
\usepackage{txfonts}
\begin{document}

\title{Symmetry-protected topological phase transitions and robust chiral order on a tunable zigzag lattice}

\author{Qibin Zheng$^{1,3}$, Xing Li$^1$}

\author{Haiyuan Zou$^2$}
\email{wlzouhy@sjtu.edu.cn}

\affiliation{$^1$Institute of Biomedical Engineering, University of Shanghai for Science and Technology, Shanghai 200093, China}
\affiliation{$^2$Tsung-Dao Lee Institute, Shanghai Jiao Tong University, Shanghai 200240, China}
\affiliation{$^3$School of Optical-Electrical and Computer Engineering, University of Shanghai for Science and Technology, Shanghai 200093, China}

\date{\today}

\begin{abstract}
Symmetry fractionalization, generating a large amount of symmetry-protected topological phases, provides scenarios for continuous phase transitions different from spontaneous symmetry breaking. However, it is hard to detect these symmetry-protected topological phase transitions experimentally. Motivated by the recent development of highly tunable ultracold polar molecules, we show that the setup in a zigzag optical lattice of this system provides a perfect platform to realize symmetry-protected topological phase transitions. By using infinite time-evolving block decimation, we obtain the phase diagram in a large parameter regions and find another scheme to realize the long-sought vector chiral phase, which is robust from quantum fluctuations. We discuss the existence of the chiral phase by an effective field analysis.
\end{abstract}

\maketitle
\section{Introduction}
\label{intro}
The last decade has witnessed significant progress in the understanding of a new type of ground states of matter, symmetry-protected topological (SPT) states~\cite{chen2013symmetry,senthil2015symmetry,chen2011complete}. Different from a traditional Ginzburg-Landau mechanism of phase characterization by symmetry breaking, SPT states arise from fractionalization of the unbroken symmetry and can be classified by the cohomology group~\cite{chen2011complete}. These states, containing only short-range entanglement, are separated from their excited states by an energy gap in the bulk but with edge or surface states protected by symmetries. The phase transitions between SPT states without symmetry breaking can be captured by crossing from one topological class to another with a gap closing. There are three possible scenarios for gap closing: with a first order critical point, a continuous quantum critical point, or an intermediate gapless phase~\cite{DHLee2015}, where the latter two cases are ``Landau forbidden" phase transitions without breaking symmetry. 

The phase transitions between interacting SPT phases have been widely studied~\cite{DHLee2017,Furukawa2012,Ueda2014Chiral2}. For a ``Landau forbidden" phase transition between interacting bosonic SPT phases, a constraint on the central charge for conformal field theories (CFTs), namely, $c\ge 1$~\cite{DHLee2017}, is obtained from studies on the $Z_n\times Z_n$ clock chain model. Another well-known example to study SPT phases is a chain model with interaction frustration beyond the nearest neighbor. In this system, a special case of an intermediate gapless phase, a vector chiral phase in between different SPT phases, is obtained in particular parameter regions (for example, with a small nearest-neighbor interaction and a large antiferromagnetic next-nearest-neighbor interaction) described by solid state materials such as copper oxides~\cite{FurukawaPRL2010}. Experimentally, few SPT phases that arise from realistic spin models can be realized by cold atom systems~\cite{bSPT2018,fSPTarxiv2018}. However, to understand the phase transitions, schemes for tunable parameters are demanded. 

Motivated by the recent experimental realization of a spin-$1/2$ model using polar molecules or magnetic atoms in optical lattices~\cite{Yan:2013xe,PhysRevLett.113.195302,PhysRevLett.107.115301,PhysRevLett.111.185305}, in this paper, we study a highly tunable one-dimensional (1D) spin-$1/2$ zigzag lattice model describing these systems. The exchange interactions can be controlled to a large unexplored parameter space by tuning the zigzag lattice structure using laser beams and tilting the dipole moment using an electric field. The frustration resulting from dipole tilting has produced many appealing signatures in previous studies on similar systems. For example, it gives rise to possible spin liquid states in a 2D setup~\cite{DSL2015,Our2017,keles2018absence,keles-prb}. By fixing the zigzag lattice to a chain of identical equilateral triangles, it covers the exactly solvable cases of two SPT phases~\cite{Our2019}. In this work, we enlarge the parameter region and explore the zigzag system with different lattice structures and find cases of all three possible scenarios of SPT phase transitions. At a particular lattice structure, a robust vector chiral phase is found. All these signatures suggest that our dipolar system serves as a perfect platform to probe the SPT phase transitions. 

The rest of this paper is organized as follows. Section~\ref{model} presents the model for dipolar spin interactions on the quasi-one-dimensional zigzag chain, and shows the tunability of the system with three parameters. We obtain the ground-state phase diagram of this model by infinite time-evolving block decimation (iTEBD) calculations and make a comparison with classical results in Sec.~\ref{phase}. Further theoretical analyses, including exact solutions of special cases and a qualitative effective field analysis, are shown in Sec.~\ref{theory}. We summarize the results in Sec.~\ref{summary}.

\section{The model}
\label{model}
We consider a spin-$1/2$ $XXZ$ model on a quasi-one-dimensional zigzag chain, 
\begin{equation}
H=\sum_{i,j}J_{i,j}[S^x_iS^x_j+S^y_iS^y_j+\eta S^z_iS^z_j],
\label{eq:model}
\end{equation}
where $\vec{S}=\vec{\sigma}/2$, and $\vec{\sigma}$ refer to the Pauli matrices. $\eta$ is the exchange anisotropy. For a special case of $\eta=0$ or 1, spin interactions in $H$ reduce to the $XY$ model type or Heisenberg model type respectively. $J_{i,j}$ is the exchange coupling between sites $i$ and $j$, and it is restricted to nearest neighbors (NN) and next nearest neighbors (NNN), with $J_{2i-1,2i}=J_1,J_{2i,2i+1}=J'_1$, and $J_{i,i+2}=J_2$.

\begin{figure}[h]
\centering
\includegraphics[width=0.4\textwidth]{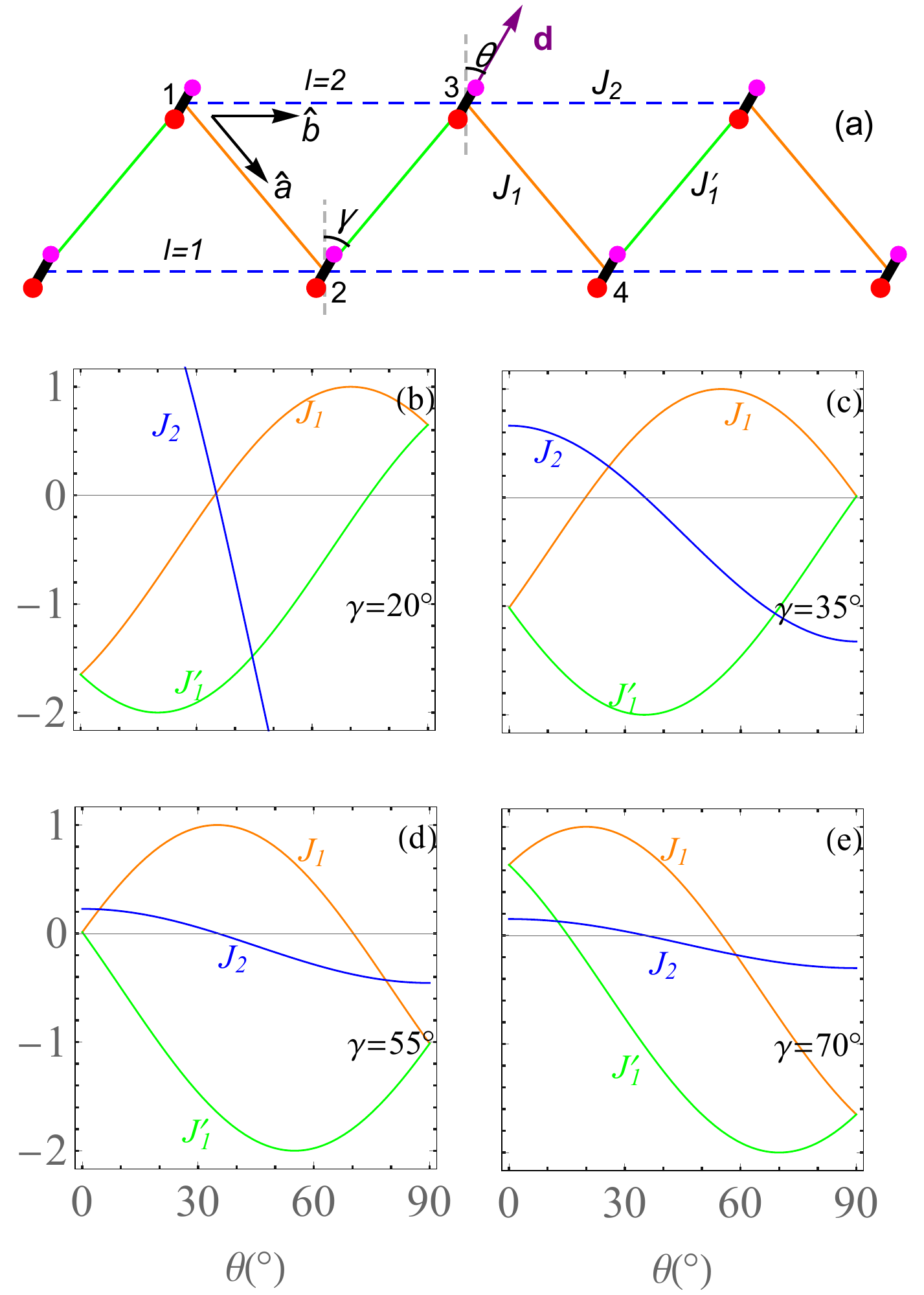}
\caption{Lattice structure and interaction couplings of the model are shown. (a) There is one dipolar molecule localized on each site of a zigzag chain. The dipoles point to the $\mathbf{d}$ direction controlled by an external electric field, forming an angle $\theta$ with the vertical direction. The exchange couplings $J_1$, $J'_1$, and $J_2$ are defined in the text. (b)-(e) The variation of the exchange couplings as functions of $\theta$ for different $\gamma$ are shown. We only list four special examples $\gamma=20^\circ, 35^\circ, 55^\circ$, and $70^\circ$ to characterize the trend of changes of couplings. At $\gamma\sim35^\circ$, the NN couplings vanish at $\theta=90^\circ$, and at $\gamma\sim55^\circ$, the NN couplings vanish at $\theta=0^\circ.$
 }
\label{fig:model}
\end{figure}

The model Eq.~\eqref{eq:model} can be realized in polar molecules such as KRb and NaK localized in deep optical lattices ~\cite{Yan:2013xe,PhysRevLett.113.195302,PhysRevLett.107.115301}. Here two rotational states of the molecules form a spin-$1/2$ system, and the exchange coupling $J_{i,j}$ refers to the dipolar interaction between two dipoles with a distance $\mathbf{r}_{ij}=\mathbf{r}_i-\mathbf{r}_j$. Formally, $J_{i,j}=J[1-3(\hat{r}_{ij}\cdot\hat{d})^2]/r_{ij}^3$ with energy unit $J>0$, and $\hat{d}$ refers to the direction of the dipoles controlled by an external electric field~\cite{DSL2015,Our2017,Our2019}. For a lattice with zigzag angle $\gamma$ (Fig.~\ref{fig:model}), applying an in-plane uniform external field on the direction with angle $\theta$ to the $y$ axis results in alternating NN exchanges and uniform NNN coupling via $J_1=J[1-3\cos^2(\theta+\gamma)]$, $J'_1=J[1-3\cos^2(\theta-\gamma)]$, and $J_2=J[1-3\sin^2\theta]/8\sin^3\gamma$. Considering large lattice spacing, longer range interactions beyond NNN are neglected.    

The anisotropy $\eta\in [0,1]$ can be tuned by varying the strength of the electric field~\cite{Yan:2013xe,DSL2015}, and the zigzag angle $\gamma\in(0,90^\circ]$ can be determined by fixing the lattice structure through laser beams. The $\gamma=30^\circ$ case, which consists of a chain of identical equilateral triangles, was studied extensively~\cite{Our2019}. In this paper, we consider all the possible cases of $\gamma$. 

By gradually tilting the electric field, so does the direction of the dipole moment $\hat{d}$ change, from perpendicular to the $J_2$ bonds to aligning with $J_2$ bonds, and the exchange coupling $J_1,J_1',J_2$ goes through a varied parameter region. Considering different $\gamma$, all possible parameter regions are covered. Figure~\ref{fig:model} shows $J_1,J_1',J_2$ as functions of $\theta$ for different $\gamma$. By increasing $\gamma$, the system changes from the limit of two chains with dominant coupling $J_2$ and a small perturbation $J_1,J_1'$ at $\gamma\sim 0$ to the limit of one chain with dominant coupling $J_1$ (or $J_1'$) and small perturbation $J_2$ at $\gamma=90^\circ$. Looking at the sign of coupling, $J_2$ changes from an antiferromagnetic (AF) to ferromagnetic (FM) case as $\theta$ changes from $0^\circ$ to $90^\circ$. However, the effects on the sign of $J_1,J_1'$ are strongly dependent on the value of $\gamma$. At small $\gamma$, $J_1,J_1'<0$ for small $\theta$ and $J_1,J_1'>0$ for large $\theta$, while at large $\gamma$, $J_1,J_1'>0$ for small $\theta$ and $J_1,J_1'<0$ for large $\theta$. There are two special cases: At $\gamma\sim 35^\circ$, $J_1=J_1'=0$ at $\theta=90^\circ$, and at $\gamma\sim 55^\circ$, $J_1=J_1'=0$ at $\theta=0^\circ$. In general, $J_1\neq J_1'$, except for $\theta=0^\circ$ and $90^\circ$.
 
Adding the contribution from the anisotropy $\eta$, the system can go through a huge parameter space with three parameters. The phase properties with the change of these parameters are then greatly desired. In the plane $\theta=0$, or cases of $J_1=J_1'$ with $J_2>0$ are extensively studied~\cite{Furukawa2012}. A vector chiral phase appears but it is very sensitive to the bond alternation. A small difference between $J_1$ and $J_1'$ will kill the chiral order~\cite{Ueda2014Chiral,Our2019}. The effect of large alternation is an interesting question. In our result, we find different vector chiral regions at $\gamma\sim 25^\circ$ and $35^\circ$, which are robust when the alternations of $J_1$ and $J_1'$ are large. The case of $J_1=J_1'\ll J_2$ is numerically hard to study~\cite{Furukawa2012}. In our system, we find a small alternation on $J_1$ and $J_1'$ will destroy the dimer phases to a Tomonaga-Luttinger liquid (TLL) phase, which means that the dimer phases at small $J_1,J_1'$ are fragile. We also show that there are plenty of exactly solvable lines for different SPT phases. We also find a large region where it is  experimentally easy to detect the phase transition between different SPT phases.

\section{Phase diagram}
\label{phase}
\begin{figure}[h]
\centering
\hspace{-1cm}\includegraphics[width=0.45\textwidth]{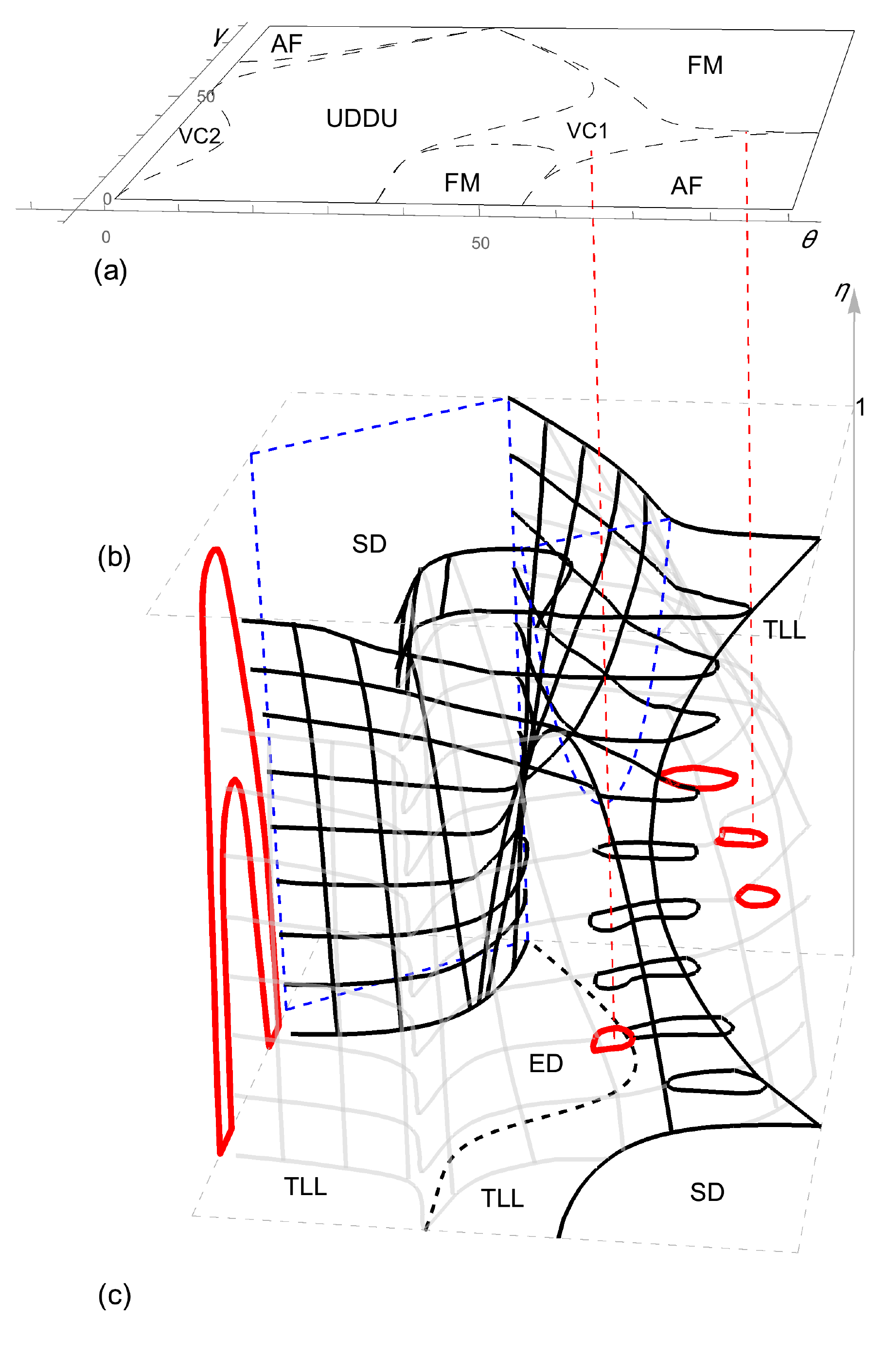}\vspace{-1cm}
\includegraphics[width=0.38\textwidth]{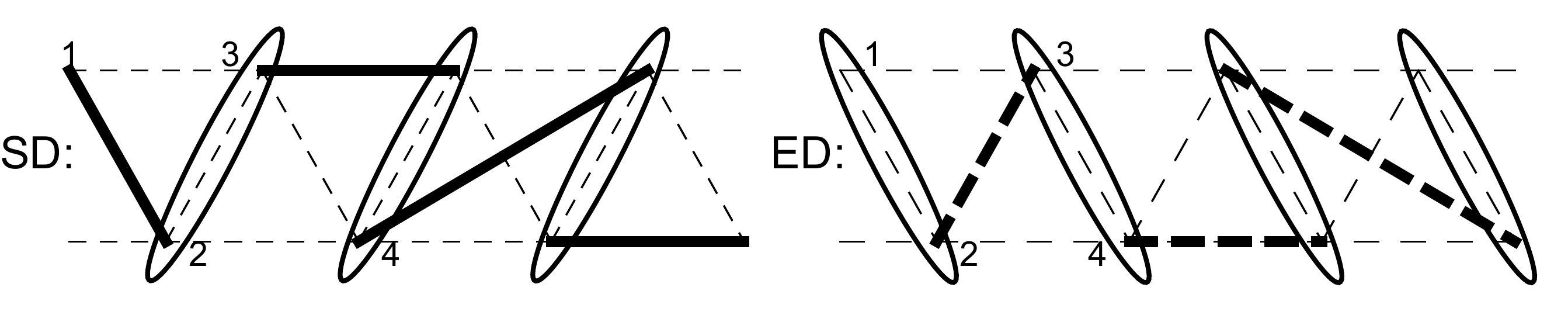}
\caption{Phase diagrams and SPT phases: (a) The classical phase diagram in the $\theta$-$\gamma$ plane. There are the AF phase, FM phase, UDDU classical dimer phase, and two regions of incommensurate vector chiral phases, the VC1 and VC2 phase. (b) The quantum phase diagram in $\theta$-$\gamma$-$\eta$ space. There are FM phases only in the $\eta=1$ plane, and two quantum dimer SPT phases: the SD phase (bounded by black grids) and ED phase (in between the black and grayed grids), gapless TLL phases, and two types of vector chiral phases (red). The first case is in the $\theta=0^\circ$ plane (or $J_1=J_1'$), corresponding to the VC2 phase in the classical phase diagram, which can be destroyed by a tiny difference between $J_1$ and $J_1'$. The second case is at $\gamma\sim25^\circ$ and $35^\circ$, corresponding to the boundary between the AF and FM phase in the VC1 phase of the classical phase diagram. The lines of exactly solvable cases are shown as two blue planes and a black dashed line for the SD and ED phases respectively. (c) The illustrations of the SD and ED phases on a zigzag system formed by thin dashed lines. One oval stands for effective spin-1’s defined in Eq. (2). A thick solid line in the SD case indicates a singlet and a thick dashed line in the ED case stands for an even-parity bond.}
\label{fig:phaseD}
\end{figure}

We first obtain a coarse classical phase diagram of the model Eq.~(\ref{eq:model}) by a Luttinger-Tisza analysis~\cite{LuttingerTisza} as a guide. Using a planar helix ansatz $\mathbf{S}_\mathbf{r}=\hat{a}\cos(\mathbf{Q}\cdot\mathbf{r})+\hat{b}\sin(\mathbf{Q}\cdot\mathbf{r})$ for spin on each site, where $\mathbf{Q}=(Q_a,Q_b)$ is the ordered wavevector, the average classical ground state energy depending on $\mathbf{Q}$ can be expressed as $E_{\textrm{cl}}=J_1\cos(Q_a)+J_1'\cos(Q_a-Q_b)+2J_2\cos(Q_b)$. Comparing different sets of wavevectors $\{\mathbf{Q}\}$ to minimize the energy, we obtain several different phases in the classical phase diagram [Fig.~\ref{fig:phaseD}(a)]: The FM phase with $\mathbf{Q}=(0,0)$, the AF phase with $\mathbf{Q}=(\pi,0)$, the spin up-down-down-up (UDDU) phase, dubbed as the classical dimer (CD) phase, with $\mathbf{Q}=(\pi, \pi)$, and an incommensurate vector chiral (VC) phases with $\mathbf{Q}\sim (x, 0)$ for VC1 and $\mathbf{Q}\sim (\pi, y)$ for VC2. All the magnetically ordered classical phases will be suppressed by quantum fluctuation. To figure out the effect of quantum fluctuation and how the phases are changed, we use the iTEBD method to extensively study the model in a large parameter space. Details on the iTEBD implementation can be found in the Appendix. The obtained phase diagram is shown in Fig.~\ref{fig:phaseD}(b). In the quantum limit, the CD and AF phase in the classical phase diagram will be dominated by two gapped SPT phases, i.e. a singlet dimer (SD) phase and an even-parity dimer (ED) phase, and a gapless TLL phase. The large region of the classical vector chiral phase has shrunk. The FM phase survives only at $\eta=1$. The two SPT phases [illustrated in Fig.~\ref{fig:phaseD}(c)] are protected by the same symmetry, e.g., any one of the following three symmetries: (i) time reversal, (ii) spatial inversion about a bond center, and (iii) $D_2$ symmetry of spin rotation~\cite{Pollmann2012}, but are topologically distinct by different edge states of systems with an open boundary condition. The properties of the two SPT phases, e.g. projective representations and properties of edge states, have been studied before~\cite{Ueda2014Chiral2,Our2019}. The significance of this work is that, besides a region of a gapless vector chiral phase near $J_1=J_1'$ (the corresponding region in the classical phase diagram is the VC2 region), which can be destroyed by a tiny alternation between $J_1$ and $J_1'$, we find additional robust regions of the VC phase with a large alternation of $J_1$ and $J_1'$ near $\gamma=25^\circ$ and $35^\circ$, where the corresponding regions in the classical phase diagram are the boundary between the FM and AF phases in the VC1 phase. 

For the 1D system, finite local magnetization has vanished and the two-spin correlation decays fast due to a strong quantum fluctuation for all the phases. Other quantities are demanded to characterize the phases and detect the phase transitions. The string order parameter~\cite{string1989}, defined as
\begin{equation}
{O}^z_n  =  - \! \lim_{r\to\infty}\langle(\hat{S}^z_{n} \! + \! \hat{S}^z_{n+1}) e^{ i \pi \! \sum_k \! \hat{S}^z_k }
(\hat{S}^z_{2r+n} \! + \! \hat{S}^z_{2r+n+1})\rangle,
\label{eq:string}
\end{equation} 
is a good quantity for this purpose, where the $k$ sum is restricted to ${n+2\leq k \leq 2r+n-1}$. A finite ${O}^z$ suggests that neighboring spins $\hat{S}^z_n+\hat{S}^z_{n+1}$ form an effective spin-1 degree of freedom and a hidden non-local order. The SD (ED) phase can be characterized by a finite $O^z_n$ value for an even (odd) site, say, $n=2$ ($n=1$). Three scenarios of phase transition, with a first-order transition point, with a continuous critical point, or with a gapless intermediate region between the SD and ED phase are all observed in the phase diagram. The first-order phase transition case is shown in the case $\gamma=30^\circ$~\cite{Our2019}. A large region of a SD-to-ED continuous phase transition at one critical point is also observed. For example, at $\gamma\ge 55^\circ$, the transition appears even at $\eta=0$. Thus, it is a perfect place for an experiment to detect the phase transition with a small electric field.
The TLL phase and VC phase are examples of the scenario when a SPT phase transition goes through a gapless intermediate region, so we focus on the latter case and use the order parameter 
\begin{equation}
\langle \hat{\kappa}^z\rangle = \frac{1}{N} \sum_i\langle (\hat{\vec{S}}_i \times \hat{ \vec{S} }_{i+1})^z \rangle,
\label{eq:chiral}
\end{equation} 
to characterized the chiral phase.

\begin{figure}[h]
\centering
\includegraphics[width=0.45\textwidth]{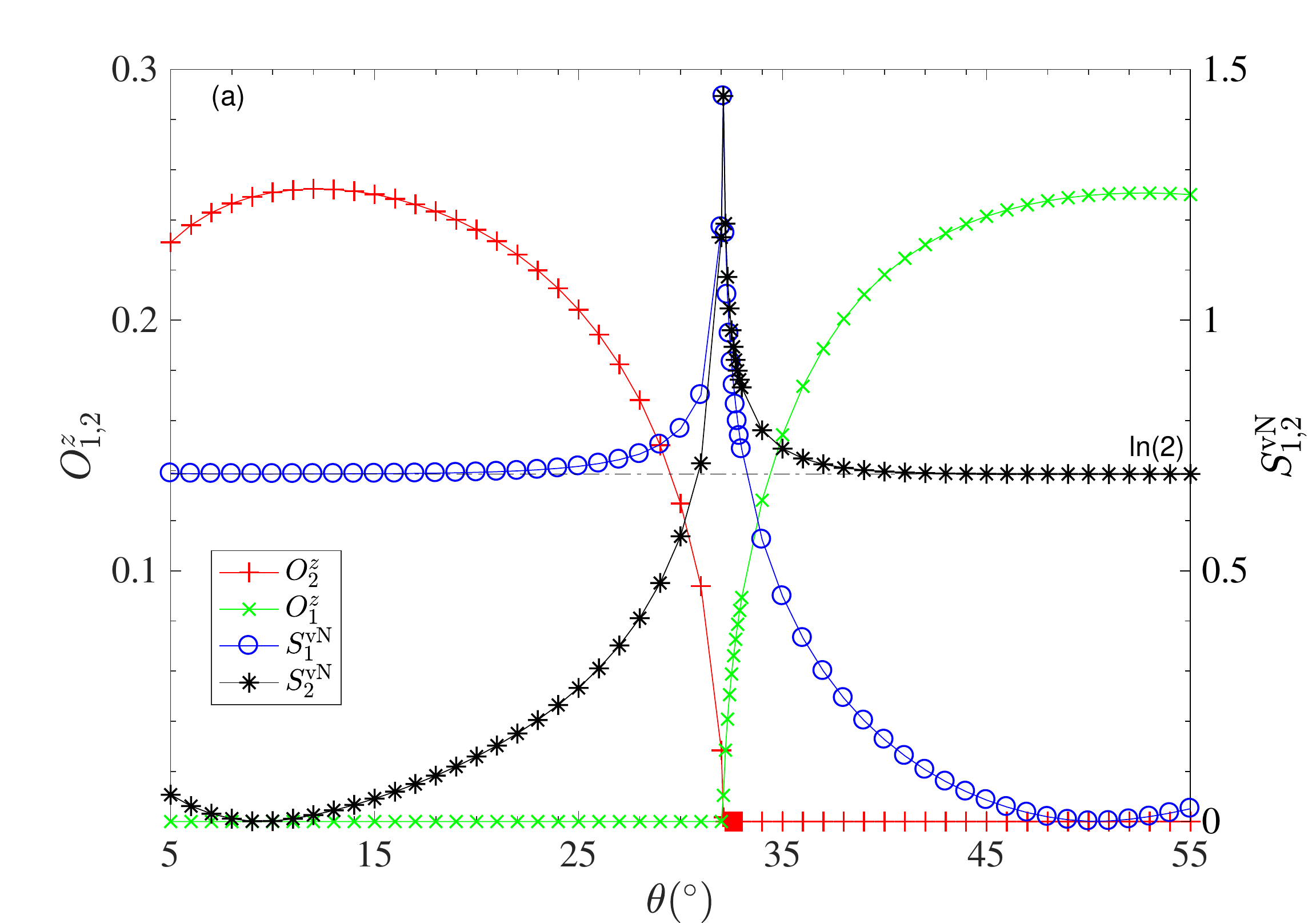}
\includegraphics[width=0.45\textwidth]{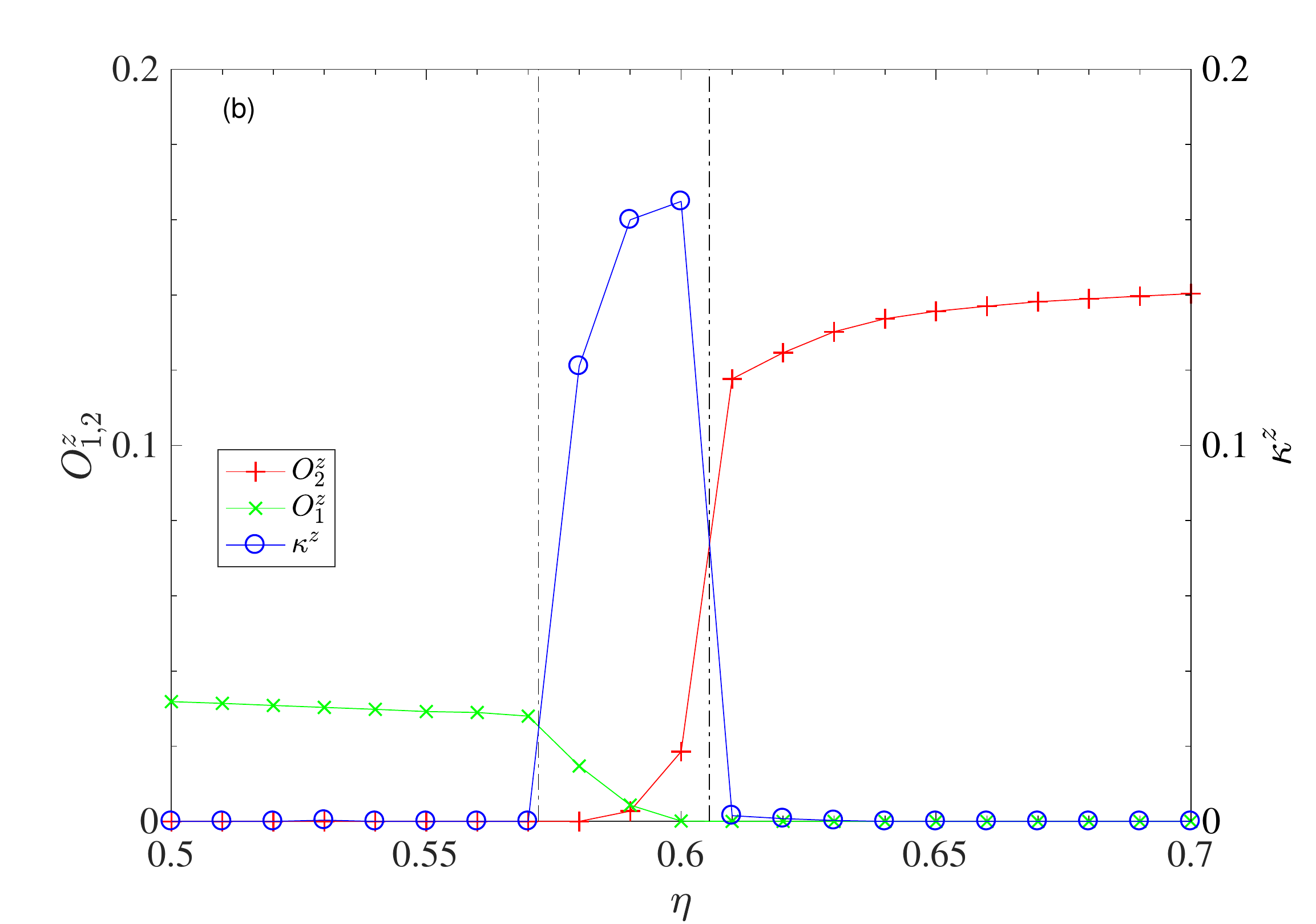}
\caption{Quantities to characterize phase transitions between two SPT phases. (a) For a direct transition with a continuous critical point case, entanglement entropy $S^{\rm vN}_{1,2}$ on a $J_1$ and $J_1'$ bond, the string order parameters $O^z_1$ and $O^z_2$ as functions of $\theta$ for $\gamma=70^\circ$ and $\eta=0$ with $\chi=100$ are shown. Zeros of $S^{\rm vN}_{2}$ (or $S^{\rm vN}_{1}$) correspond to an exact singlet product state (or even-parity product state) and $S^{\rm vN}_{1,2}=\ln2$ for the two cases respectively. (b) For the case going through a gapless intermediate phase, the string order parameters $O^z_1$ and $O^z_2$ and vector chirality $\kappa^z$ as functions of $\eta$ at $\theta=70^\circ$ and $\gamma=35^\circ$ with $\chi=100$ are shown. A finite region of the VC phase can be observed.}
\label{fig:Heig}
\end{figure}

Another quantity characterizing the SPT phase transition is the von Neumann entanglement entropy, which cuts the $J_1$ bond or $J_1'$ bond. Based on the matrix product state (MPS) representation of a many-body wave function, the iTEBD algorithm is a powerful method to evaluate local or nonlocal physical operators in the thermodynamic limit by considering transitional invariant of the building blocks of the wave function. The local MPSs represent local bulk wave functions and the bond vectors represent the environment and the entanglement between different bulks. The von Neumann entanglement entropy is defined through the normalized eigenvalues $\lambda_{b,\chi}$ of the bond vectors with a Schmidt rank $\chi$ by cutting a $b$ bond, with $b=1$ (or 2) for the $J_1$ (or $J_1'$) bond,
\begin{equation}
S^{\rm vN}_b=-\sum_{\chi} \lambda^2_{b,\chi}\ln\lambda^2_{b,\chi},
\label{eq:entangle}
\end{equation}
As shown in Fig.~\ref{fig:Heig}(a), jumps of the $O^z_n$ and peaks of the entanglement at a single point characterize the continuous phase transition with one critical point. The zeros of entanglement entropy by cutting either bond ($J_1$ or $J_1'$) suggest an exactly solvable point~\cite{Our2019}. One gives a singlet product state and the other gives an even-parity product state. At $\gamma\sim70^\circ$, and $\eta\sim0$, by tuning $\theta$, the system can go through the two exactly solvable cases. This suggest a perfect parameter region for experiments with a small electric field. In Fig.~\ref{fig:Heig}(b), at $\gamma\sim35^\circ$, a region with a finite chiral order parameter exists, which suggests a robust VC phase.

\section{Theoretical Analysis}
\label{theory}
\subsection{Exact solutions}
\label{exact}
It is well known that the Majumdar-Ghosh (MG) point, when $J_1=J_1'=2J_2$, for the antiferromagnet Heisenberg chain is exactly solvable. The exact ground state is a direct product of singlet dimers with twofold degeneracy~\cite{Majumdar1969}. There are many extensions
of this type of exactly solvable model~\cite{Shastry1981,Kanter}. In previous work that focused on $\gamma=30^\circ$~\cite{Our2019}, we have proved a similar exactly solvable region for both SD and ED cases. By considering all the possible $\gamma$, more exactly solvable regions are shown in Fig.~\ref{fig:phaseD}. For the SD cases, at $J_1'=2J_2>0$ with $\eta\in [0,1]$, and $J_1'=2J_2<0$ with $\eta\in [(|J_1'|-J_1)/J_1,1]$, the ground state is the product state of the spin singlet  $(\vert\uparrow\downarrow\rangle-\vert\downarrow\uparrow\rangle)/\sqrt{2}$ on all $J_1$ bonds (these regions are shown in Fig.~\ref{fig:phaseD} with a blue plane). For the ED cases, at $J_1=-2J_2$, $\eta=0$, the ground state is the product of a spin triplet  $(\vert\uparrow\downarrow\rangle+\vert\downarrow\uparrow\rangle)/\sqrt{2}$ on all $J_1'$ bonds, shown as a black dashed line in the plane $\eta=0$ in Fig.~\ref{fig:phaseD}. 

\subsection{Effective field theory}
\label{eft}
Regions of a vector chiral phase at $\gamma\sim 35^\circ$ and $\gamma\sim 25^\circ$ can be qualitatively understood by bosonization. In the limit $J_1,J_1'\ll |J_2|$, the model Eq.~(1) can be viewed as two weakly coupled ferromagnetic XXZ chains, each being a Tomonaga-Luttinger liquid. Using Abelian bosonization~\cite{Giamarchi2003}, each decoupled chain can be described by the Hamiltonian density,
\begin{equation}
\mathcal{H}_l=uK(\partial_x\theta_l)^2+u/K(\partial_x\phi_l)^2,
\end{equation}
where $l=1,2$ is the chain index and $x$ is the site index on each chain. The Luttinger parameter $K$ and velocity $u$ are given by $K={\pi}/{(2\arccos\eta)}$, $u=|J_2| \sin(\pi/2K){K}/({2K-1})$. The bosonic fields $\phi_l(x)$ and their conjugates $\partial_x\theta_l(x)/\pi$ satisfy the commutation relation $[\partial_x\theta_l(x)/\pi, \phi_l(x')]=-i\delta(x-x')$.

The spin operators in terms of the bosonic fields are
\begin{eqnarray}
S^z_l(x) &\sim& (-1)^l[\frac{1}{\pi}\Delta\phi_l(x)+\frac{(-1)^x}{\pi}\cos 2\phi_l], \nonumber\\
S^+(x) &\sim& (-1)^le^{-i\theta_l}\frac{1}{\sqrt{2\pi}}[1+(-1)^x\cos 2\phi_l]. \nonumber 
\end{eqnarray}

One can construct fields $\phi_\pm=(\phi_1\pm\phi_2)/\sqrt{2}$ and similarly $\theta_\pm$ for the symmetric ($+$) and antisymmetric ($-$) parts, and the low-energy effective Hamiltonian density takes the form $\mathcal{H}=\mathcal{H}_++\mathcal{H}_-+\mathcal{H}_{\rm{int}}$, with
\begin{eqnarray}
\mathcal{H}_+&=&u_+K_+(\partial_x\theta_+)^2+\frac{u_+}{K_+}(\partial_x\phi_+)^2+g_1\cos(\sqrt{8}\phi_+),\nonumber \\
\mathcal{H}_-&=&u_-K_-(\partial_x\theta_-)^2+\frac{u_-}{K_-}(\partial_x\phi_-)^2+g_1\cos(\sqrt{8}\phi_-)\nonumber\\
&+&g_2\cos(\sqrt{2}\theta_-),\nonumber \\
\mathcal{H}_{\rm{int}}&=&g_3\cos(\sqrt{2}\theta_-)\cos(\sqrt{8}\phi_+)+\partial_x\theta_{\pm} \rm{terms}...,
\label{eq:effect}
\end{eqnarray}
where the coupling constants $g_1\sim (J_1-J_1')\eta/\pi$, $g_2\sim J_1+J_1'$, $g_3\sim\ ({J_1-J_1'})/{2}$, and $u_\pm=u \beta_\pm$, $K_\pm=K/\beta_\pm$, with $\beta_\pm= [1\pm {K(J_1+J_1')\eta}/(\pi u)]^{1/2}$ and the term $\partial_x\theta_{\pm}$ corresponds to the vector chiral operators.

The behavior of $\mathcal{H}$ is described by the renormalization group (RG) equations for the effective couplings,
\begin{equation}
\label{eq:rg}
\frac{dg_1(l)}{dl}=(2-2K_+)g_1(l), \frac{dg_2(l)}{dl}=[2-1/(2K_-)]g_2(l),
\end{equation}
where  $l=-\ln(\Lambda/\Lambda_0)$ and $\Lambda$ is the cutoff of RG steps.

In the RG equation Eq.~\ref{eq:rg}, $2-1/(2K_-)$ is always positive, and the term $\cos(\sqrt{2}\theta_-)$ in $\mathcal{H}_-$ is relevant. In a mean-field treatment, one can replace $\cos(\sqrt{2}\theta_-)$ with $\Delta=\langle\cos(\sqrt{2}\theta_-)\rangle$ in $\mathcal{H}_{\rm{int}}$ and combine it with the $\cos(\sqrt{8}\phi_+)$ term in $\mathcal{H}_+$, with $g_1\rightarrow g_1+g_3\Delta$ to make the dimer operator more relevant. However, at $\gamma\sim 35^\circ, \theta\sim 80^\circ$, $J_1\sim -J_1'$, ($g_2\sim 0$), $\langle\cos(\sqrt{2}\theta_-)\rangle$ can not be condensed. Thus, it can not contribute to the $g_1$ term in $\mathcal{H}_+$ to make $\cos(\sqrt{8}\phi_+)$ more relevant. The chiral term with $\partial_x\theta_{\pm}$ in $\mathcal{H}_{\rm{int}}$ can be comparable with the $g_1$ term in the sine-Gordon Hamiltonian $\mathcal{H}_+$, which can produce a relevant vector chiral order. 
\section{Summary}
\label{summary}
In summary, we have shown that the zigzag XXZ model inspired by molecular gas experiments provides a promising platform for detecting SPT phase transitions for spin-1/2 systems. This setup can realize all the scenarios of phase transition between different SPT phases. Large robust vector chiral phases against quantum fluctuation can be realized. The dynamics of the SPT phase transitions deserve future investigations.

\section{Acknowledgments}
H.Z. thank Erhai Zhao for helpful discussions. This work is supported by National Natural Science Foundation of China Grant No. 11804221 (H. Z.), and Science and Technology Commission of Shanghai Municipality Grant No. 16DZ2260200 (H. Z.).

\appendix*
\section{iTEBD calculation}
\begin{figure}[h]
\centering
\includegraphics[width=0.45\textwidth]{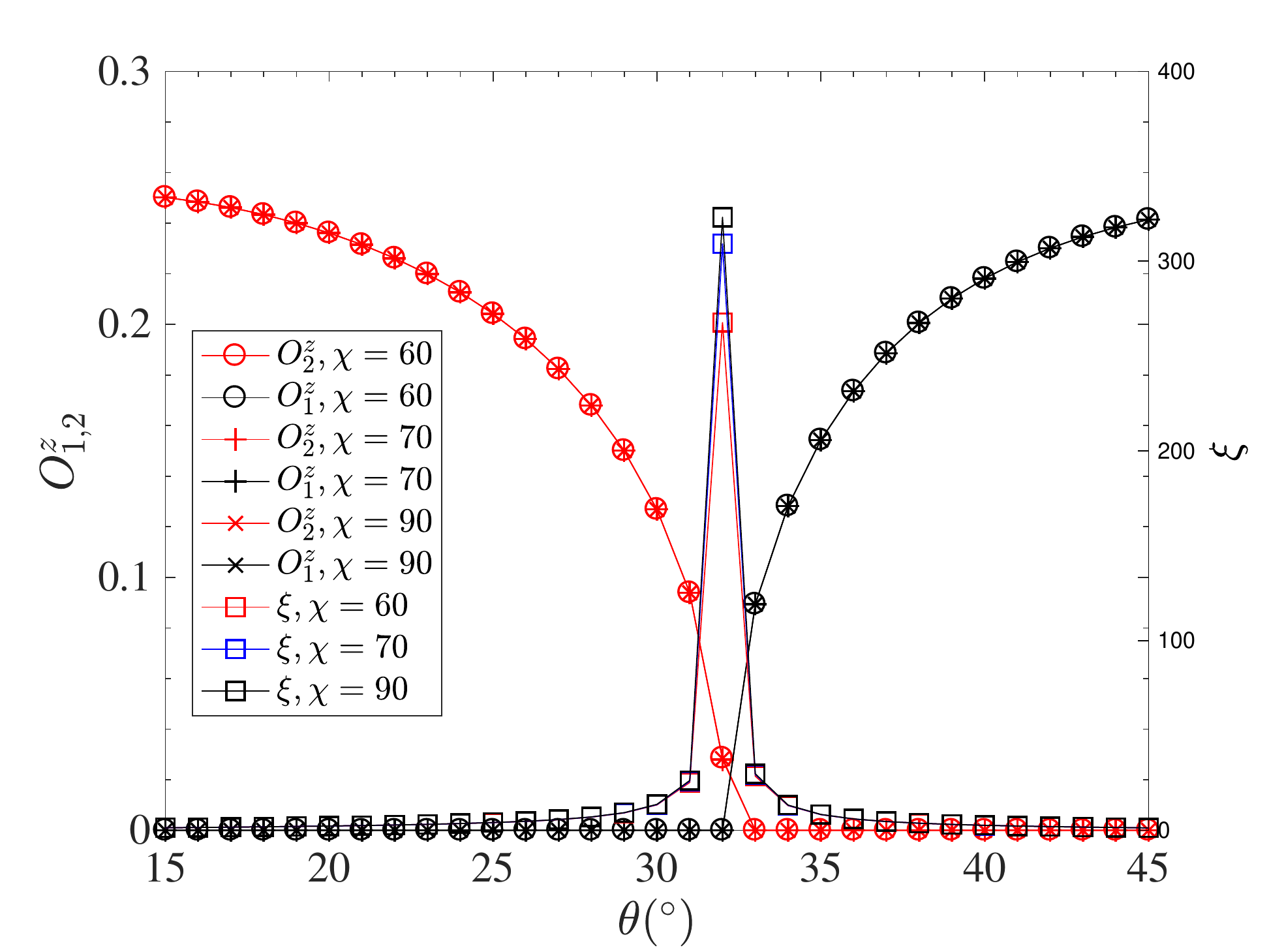}
\caption{String order parameters and correlation length for $\gamma=70^\circ$ and $\eta=0$ with bond dimensions $\chi=60$,70, and 90.}
\label{fig:Dbond}
\end{figure}

We use a four-tensor unit cell with a virtual bond dimension (Schmidt rank) $\chi$ to form the initial matrix product state (MPS) in the iTEBD calculation. Imaginary time evolution with the time interval $dt = 0.02J^{-1}$ is performed until the ground state is convergent. The string order parameter and entanglement entropy defined in the main text are calculated from the obtained ground state. By scanning the 3D parameter space $(\gamma-\theta-\eta)$ with $\gamma\in[0,90^\circ]$, $\theta\in[0,90^\circ]$, $\eta\in[0,1]$, and increasing the Schmidt rank $\chi$ upto 100, convergent results of the order parameters can be reached at most regions of the parameter space except for the SD-to-TLL transition regions, where $\chi=300$ is needed. To illustrate the convergence of results at SD-to-ED transition area with $\chi<100$, we also calculate the spin-spin correlation, from which the correlation length $\xi$ can be obtained by the relation~\cite{Furukawa2012}:
\begin{equation}
\langle\mathbf{S_1}\cdot\mathbf{S_{1+r}}\rangle\sim r^{-1/2}e^{-r/\xi}
\end{equation}
 Figure~\ref{fig:Dbond} shows the string order parameters and correlation lengths calculated at different bond dimensions $\chi$ at $\gamma=70^\circ$ and $\eta=0$. Both the overlapping of string order parameters and only a tiny increase of correlation lengths near the phase transition point with increased $\chi$ suggest that the results are convergent.

\begin{figure*}
\includegraphics[width=0.45\textwidth]{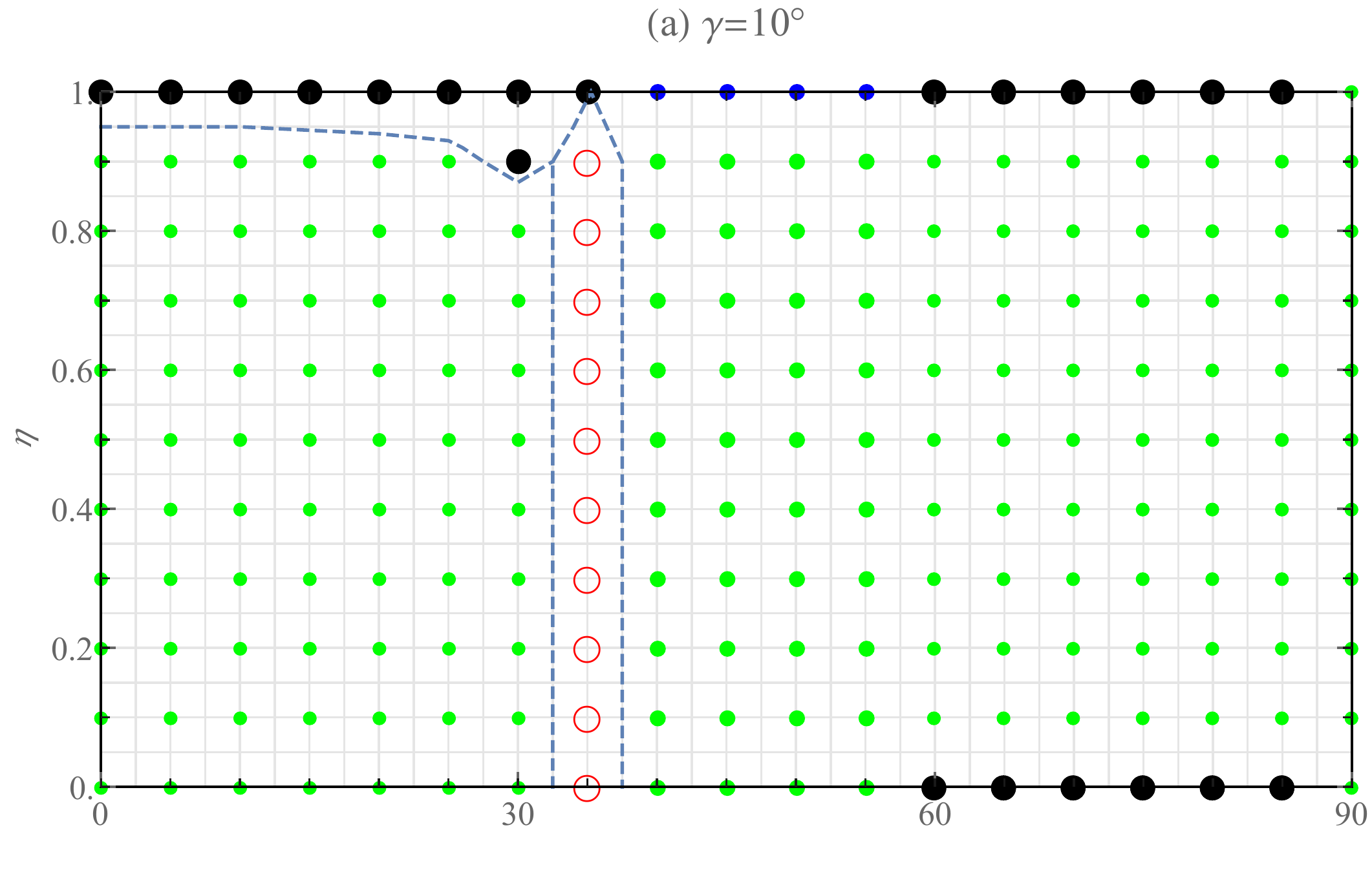}
\includegraphics[width=0.45\textwidth]{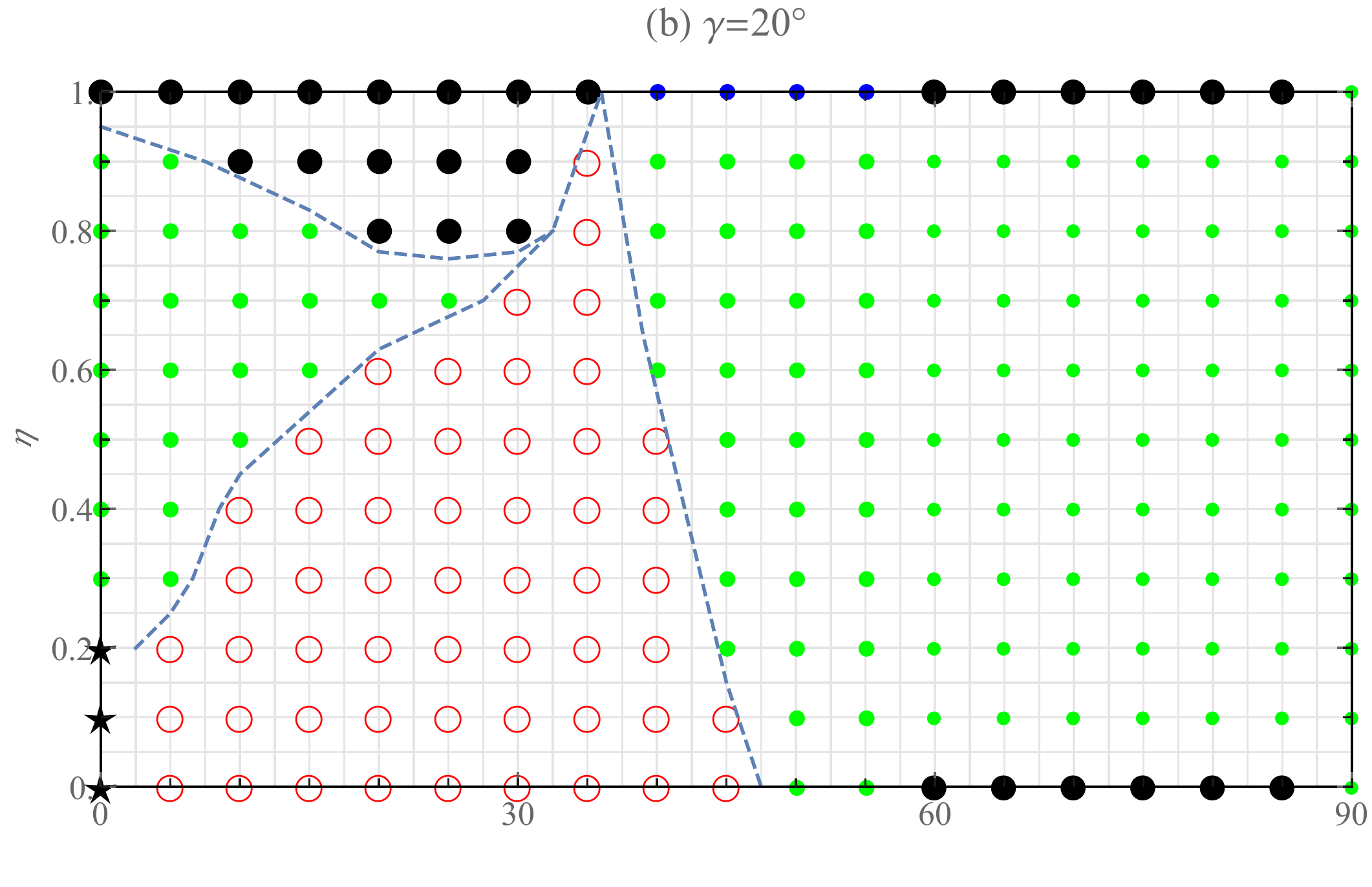}
\includegraphics[width=0.45\textwidth]{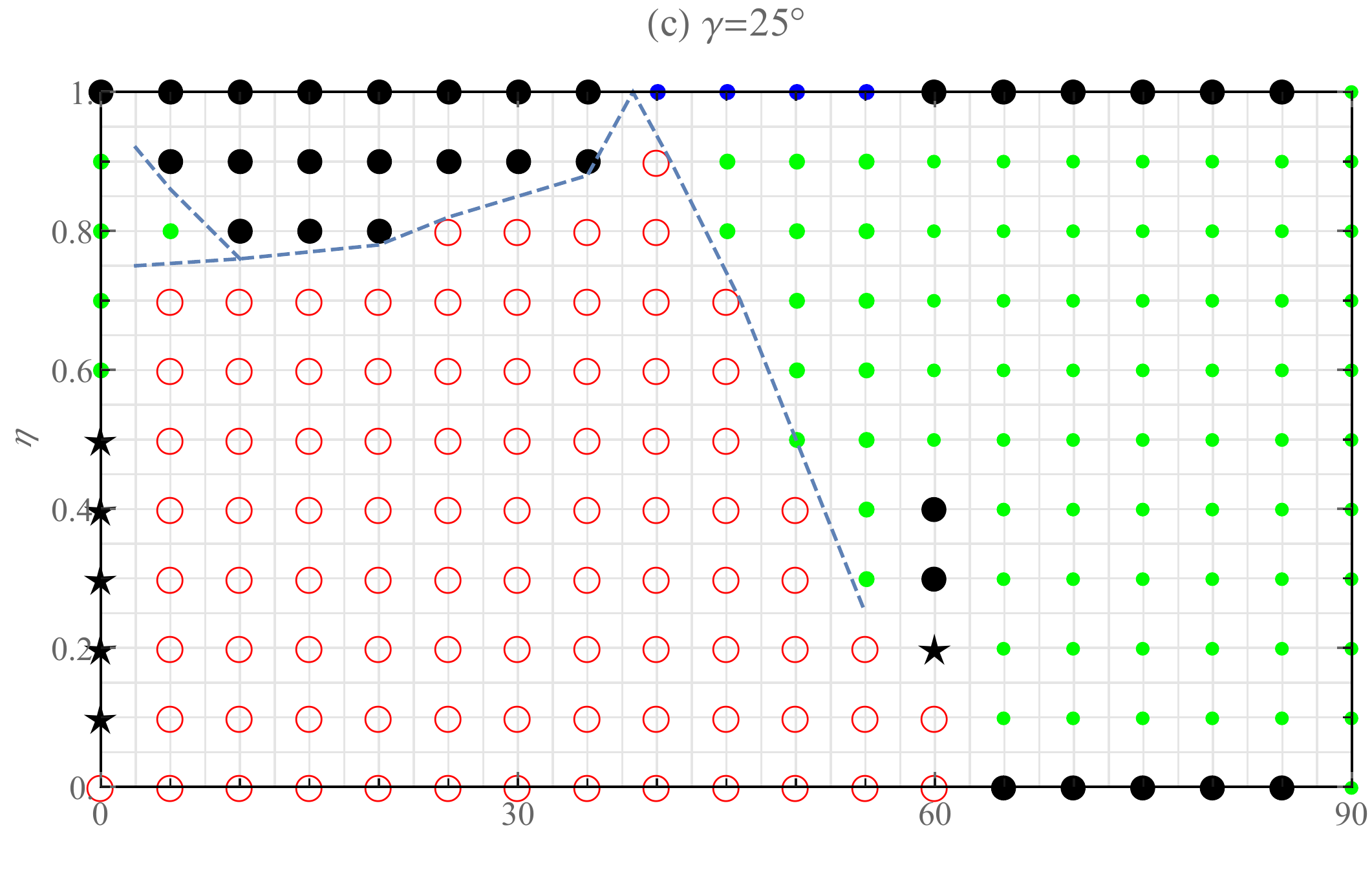}
\includegraphics[width=0.45\textwidth]{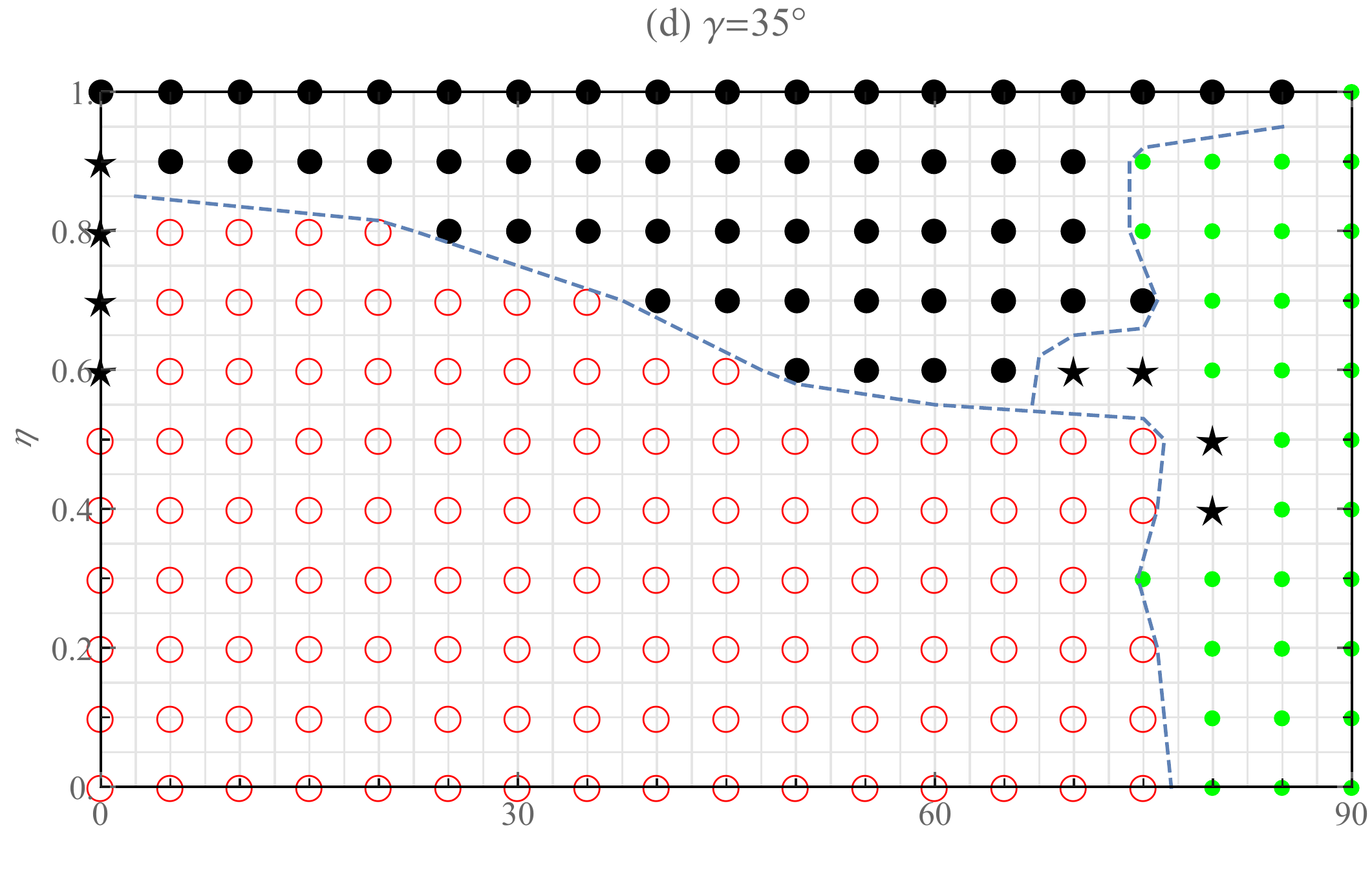}
\includegraphics[width=0.45\textwidth]{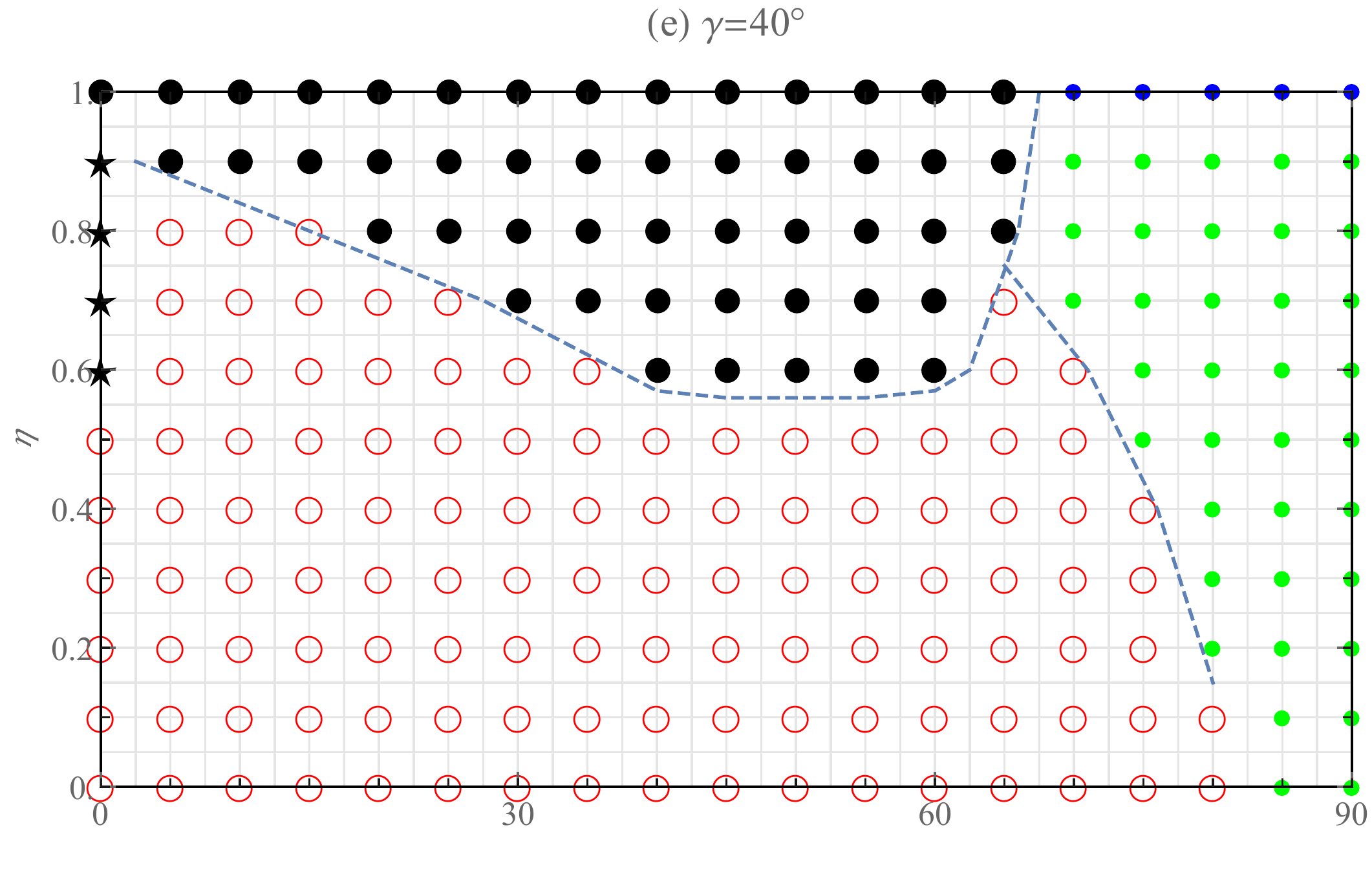}
\includegraphics[width=0.45\textwidth]{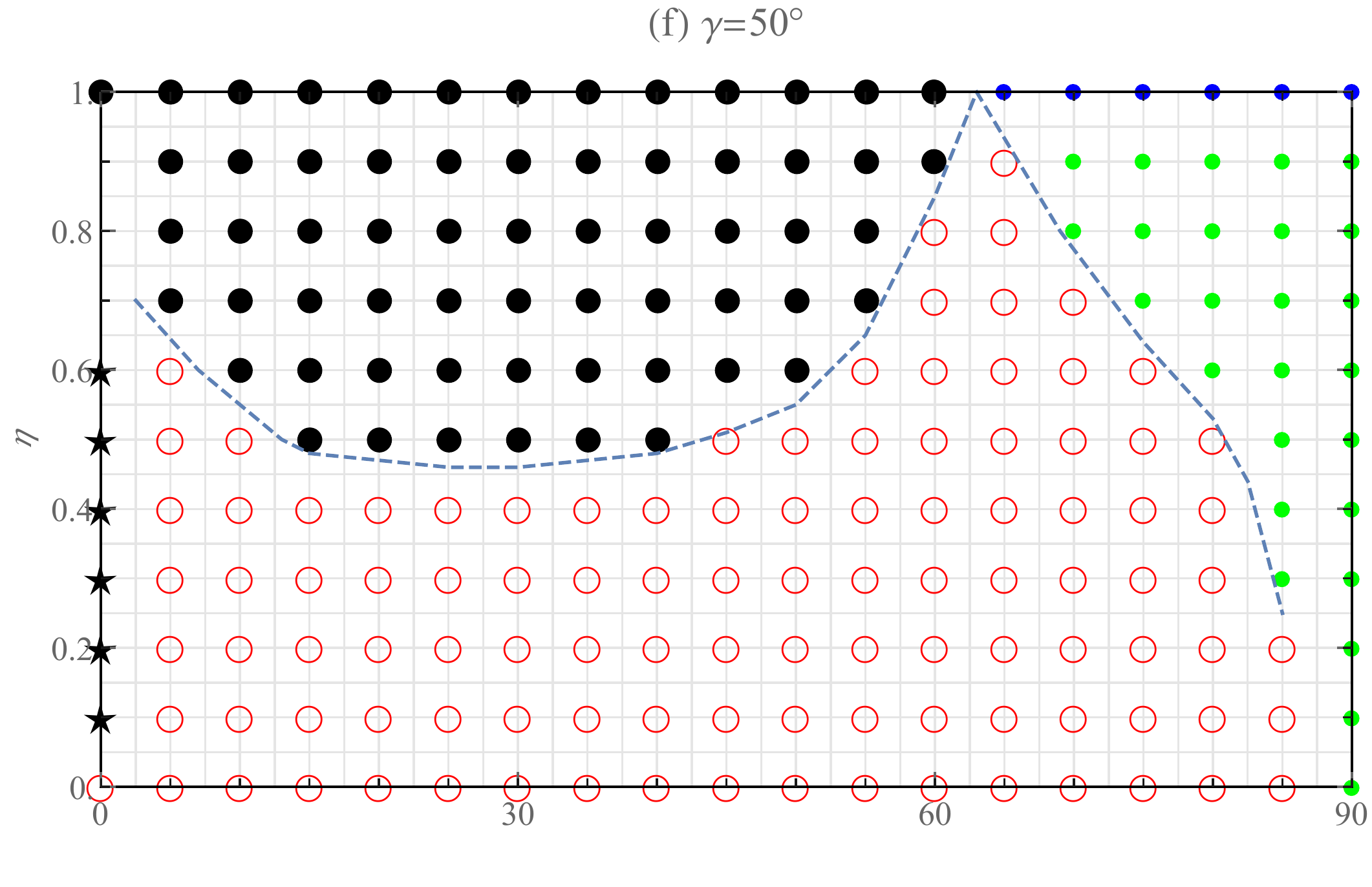}
\includegraphics[width=0.45\textwidth]{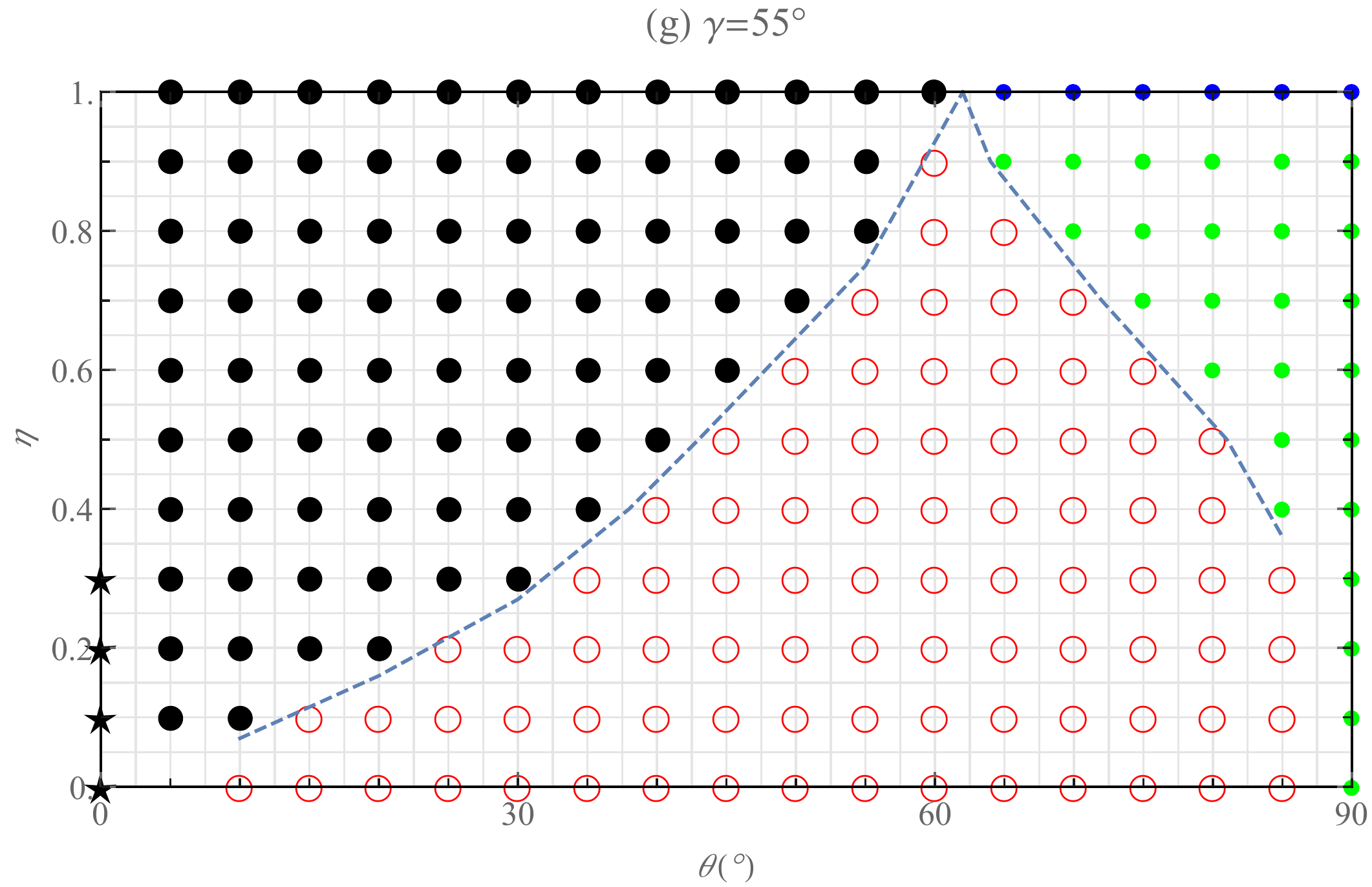}
\includegraphics[width=0.45\textwidth]{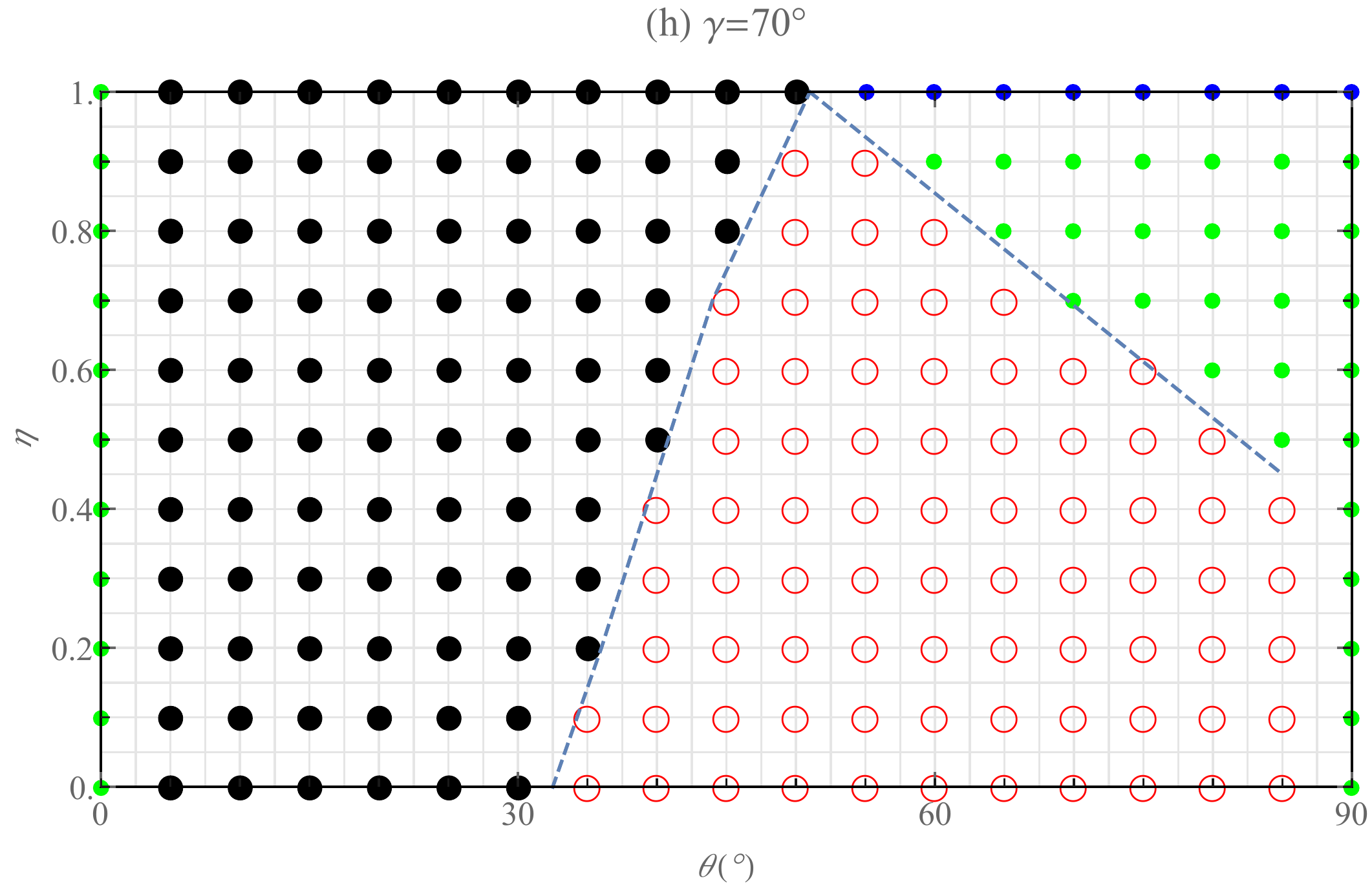}
\caption{2D phase diagrams in the $(\theta$-$\eta)$ plane with different $\gamma$ (a)-(h). The ED phase is indicated by an open circle (\textcolor{red}{$\medcirc$}), the SD phase is indicated by a solid black circle ($\medbullet$), the TLL phase is indicated by a solid green circle  (\textcolor{green}{$\bullet$}), the FM phase is indicated by a solid blue circle (\textcolor{blue}{$\bullet$}),and the chiral phase is indicated by a solid black star ($\bigstar$). The dashed line which separates different phases is a guide to the eye.}
\label{fig:2dphase1}
\end{figure*}

We first obtained slices of 2D phase diagrams on the $(\theta$-$\eta)$ plane with fixed $\gamma$ (Fig.~\ref{fig:2dphase1} shows examples of 2D phase diagrams with different $\gamma$). The 3D phase diagram (Fig.2 in the main text) is then obtained by combining all 2D $\gamma$-fixed slices with error bars totally determined by the parameter intervals, i.e., $d\gamma=10^\circ$ ($5^\circ$ near $\gamma\sim30^\circ$), $d\theta=5^\circ$, and $d\eta=0.1$.


\begin{thebibliography}{27}%
\makeatletter
\providecommand \@ifxundefined [1]{%
 \@ifx{#1\undefined}
}%
\providecommand \@ifnum [1]{%
 \ifnum #1\expandafter \@firstoftwo
 \else \expandafter \@secondoftwo
 \fi
}%
\providecommand \@ifx [1]{%
 \ifx #1\expandafter \@firstoftwo
 \else \expandafter \@secondoftwo
 \fi
}%
\providecommand \natexlab [1]{#1}%
\providecommand \enquote  [1]{``#1''}%
\providecommand \bibnamefont  [1]{#1}%
\providecommand \bibfnamefont [1]{#1}%
\providecommand \citenamefont [1]{#1}%
\providecommand \href@noop [0]{\@secondoftwo}%
\providecommand \href [0]{\begingroup \@sanitize@url \@href}%
\providecommand \@href[1]{\@@startlink{#1}\@@href}%
\providecommand \@@href[1]{\endgroup#1\@@endlink}%
\providecommand \@sanitize@url [0]{\catcode `\\12\catcode `\$12\catcode
  `\&12\catcode `\#12\catcode `\^12\catcode `\_12\catcode `\%12\relax}%
\providecommand \@@startlink[1]{}%
\providecommand \@@endlink[0]{}%
\providecommand \url  [0]{\begingroup\@sanitize@url \@url }%
\providecommand \@url [1]{\endgroup\@href {#1}{\urlprefix }}%
\providecommand \urlprefix  [0]{URL }%
\providecommand \Eprint [0]{\href }%
\providecommand \doibase [0]{http://dx.doi.org/}%
\providecommand \selectlanguage [0]{\@gobble}%
\providecommand \bibinfo  [0]{\@secondoftwo}%
\providecommand \bibfield  [0]{\@secondoftwo}%
\providecommand \translation [1]{[#1]}%
\providecommand \BibitemOpen [0]{}%
\providecommand \bibitemStop [0]{}%
\providecommand \bibitemNoStop [0]{.\EOS\space}%
\providecommand \EOS [0]{\spacefactor3000\relax}%
\providecommand \BibitemShut  [1]{\csname bibitem#1\endcsname}%
\let\auto@bib@innerbib\@empty
\bibitem [{\citenamefont {Chen}\ \emph {et~al.}(2013)\citenamefont {Chen},
  \citenamefont {Gu}, \citenamefont {Liu},\ and\ \citenamefont
  {Wen}}]{chen2013symmetry}%
  \BibitemOpen
  \bibfield  {author} {\bibinfo {author} {\bibfnamefont {X.}~\bibnamefont
  {Chen}}, \bibinfo {author} {\bibfnamefont {Z.-C.}\ \bibnamefont {Gu}},
  \bibinfo {author} {\bibfnamefont {Z.-X.}\ \bibnamefont {Liu}}, \ and\
  \bibinfo {author} {\bibfnamefont {X.-G.}\ \bibnamefont {Wen}},\ }\href@noop
  {} {\bibfield  {journal} {\bibinfo  {journal} {Physical Review B}\ }\textbf
  {\bibinfo {volume} {87}},\ \bibinfo {pages} {155114} (\bibinfo {year}
  {2013})}\BibitemShut {NoStop}%
\bibitem [{\citenamefont {Senthil}(2015)}]{senthil2015symmetry}%
  \BibitemOpen
  \bibfield  {author} {\bibinfo {author} {\bibfnamefont {T.}~\bibnamefont
  {Senthil}},\ }\href@noop {} {\bibfield  {journal} {\bibinfo  {journal} {Annu.
  Rev. Condens. Matter Phys.}\ }\textbf {\bibinfo {volume} {6}},\ \bibinfo
  {pages} {299} (\bibinfo {year} {2015})}\BibitemShut {NoStop}%
\bibitem [{\citenamefont {Chen}\ \emph {et~al.}(2011)\citenamefont {Chen},
  \citenamefont {Gu},\ and\ \citenamefont {Wen}}]{chen2011complete}%
  \BibitemOpen
  \bibfield  {author} {\bibinfo {author} {\bibfnamefont {X.}~\bibnamefont
  {Chen}}, \bibinfo {author} {\bibfnamefont {Z.-C.}\ \bibnamefont {Gu}}, \ and\
  \bibinfo {author} {\bibfnamefont {X.-G.}\ \bibnamefont {Wen}},\ }\href@noop
  {} {\bibfield  {journal} {\bibinfo  {journal} {Physical Review B}\ }\textbf
  {\bibinfo {volume} {84}},\ \bibinfo {pages} {235128} (\bibinfo {year}
  {2011})}\BibitemShut {NoStop}%
\bibitem [{\citenamefont {Tsui}\ \emph {et~al.}(2015)\citenamefont {Tsui},
  \citenamefont {Jiang}, \citenamefont {Lu},\ and\ \citenamefont
  {Lee}}]{DHLee2015}%
  \BibitemOpen
  \bibfield  {author} {\bibinfo {author} {\bibfnamefont {L.}~\bibnamefont
  {Tsui}}, \bibinfo {author} {\bibfnamefont {H.-C.}\ \bibnamefont {Jiang}},
  \bibinfo {author} {\bibfnamefont {Y.-M.}\ \bibnamefont {Lu}}, \ and\ \bibinfo
  {author} {\bibfnamefont {D.-H.}\ \bibnamefont {Lee}},\ }\href {\doibase
  https://doi.org/10.1016/j.nuclphysb.2015.04.020} {\bibfield  {journal}
  {\bibinfo  {journal} {Nuclear Physics B}\ }\textbf {\bibinfo {volume}
  {896}},\ \bibinfo {pages} {330 } (\bibinfo {year} {2015})}\BibitemShut
  {NoStop}%
\bibitem [{\citenamefont {Tsui}\ \emph {et~al.}(2017)\citenamefont {Tsui},
  \citenamefont {Huang}, \citenamefont {Jiang},\ and\ \citenamefont
  {Lee}}]{DHLee2017}%
  \BibitemOpen
  \bibfield  {author} {\bibinfo {author} {\bibfnamefont {L.}~\bibnamefont
  {Tsui}}, \bibinfo {author} {\bibfnamefont {Y.-T.}\ \bibnamefont {Huang}},
  \bibinfo {author} {\bibfnamefont {H.-C.}\ \bibnamefont {Jiang}}, \ and\
  \bibinfo {author} {\bibfnamefont {D.-H.}\ \bibnamefont {Lee}},\ }\href
  {\doibase https://doi.org/10.1016/j.nuclphysb.2017.03.021} {\bibfield
  {journal} {\bibinfo  {journal} {Nuclear Physics B}\ }\textbf {\bibinfo
  {volume} {919}},\ \bibinfo {pages} {470 } (\bibinfo {year}
  {2017})}\BibitemShut {NoStop}%
\bibitem [{\citenamefont {Furukawa}\ \emph {et~al.}(2012)\citenamefont
  {Furukawa}, \citenamefont {Sato}, \citenamefont {Onoda},\ and\ \citenamefont
  {Furusaki}}]{Furukawa2012}%
  \BibitemOpen
  \bibfield  {author} {\bibinfo {author} {\bibfnamefont {S.}~\bibnamefont
  {Furukawa}}, \bibinfo {author} {\bibfnamefont {M.}~\bibnamefont {Sato}},
  \bibinfo {author} {\bibfnamefont {S.}~\bibnamefont {Onoda}}, \ and\ \bibinfo
  {author} {\bibfnamefont {A.}~\bibnamefont {Furusaki}},\ }\href {\doibase
  10.1103/PhysRevB.86.094417} {\bibfield  {journal} {\bibinfo  {journal} {Phys.
  Rev. B}\ }\textbf {\bibinfo {volume} {86}},\ \bibinfo {pages} {094417}
  (\bibinfo {year} {2012})}\BibitemShut {NoStop}%
\bibitem [{\citenamefont {Ueda}\ and\ \citenamefont
  {Onoda}(2014{\natexlab{a}})}]{Ueda2014Chiral2}%
  \BibitemOpen
  \bibfield  {author} {\bibinfo {author} {\bibfnamefont {H.}~\bibnamefont
  {Ueda}}\ and\ \bibinfo {author} {\bibfnamefont {S.}~\bibnamefont {Onoda}},\
  }\href {\doibase 10.1103/PhysRevB.90.214425} {\bibfield  {journal} {\bibinfo
  {journal} {Phys. Rev. B}\ }\textbf {\bibinfo {volume} {90}},\ \bibinfo
  {pages} {214425} (\bibinfo {year} {2014}{\natexlab{a}})}\BibitemShut
  {NoStop}%
\bibitem [{\citenamefont {Furukawa}\ \emph {et~al.}(2010)\citenamefont
  {Furukawa}, \citenamefont {Sato},\ and\ \citenamefont
  {Onoda}}]{FurukawaPRL2010}%
  \BibitemOpen
  \bibfield  {author} {\bibinfo {author} {\bibfnamefont {S.}~\bibnamefont
  {Furukawa}}, \bibinfo {author} {\bibfnamefont {M.}~\bibnamefont {Sato}}, \
  and\ \bibinfo {author} {\bibfnamefont {S.}~\bibnamefont {Onoda}},\ }\href
  {\doibase 10.1103/PhysRevLett.105.257205} {\bibfield  {journal} {\bibinfo
  {journal} {Phys. Rev. Lett.}\ }\textbf {\bibinfo {volume} {105}},\ \bibinfo
  {pages} {257205} (\bibinfo {year} {2010})}\BibitemShut {NoStop}%
\bibitem [{\citenamefont {Song}\ \emph {et~al.}(2018)\citenamefont {Song},
  \citenamefont {Zhang}, \citenamefont {He}, \citenamefont {Poon},
  \citenamefont {Hajiyev}, \citenamefont {Zhang}, \citenamefont {Liu},\ and\
  \citenamefont {Jo}}]{bSPT2018}%
  \BibitemOpen
  \bibfield  {author} {\bibinfo {author} {\bibfnamefont {B.}~\bibnamefont
  {Song}}, \bibinfo {author} {\bibfnamefont {L.}~\bibnamefont {Zhang}},
  \bibinfo {author} {\bibfnamefont {C.}~\bibnamefont {He}}, \bibinfo {author}
  {\bibfnamefont {T.~F.~J.}\ \bibnamefont {Poon}}, \bibinfo {author}
  {\bibfnamefont {E.}~\bibnamefont {Hajiyev}}, \bibinfo {author} {\bibfnamefont
  {S.}~\bibnamefont {Zhang}}, \bibinfo {author} {\bibfnamefont {X.-J.}\
  \bibnamefont {Liu}}, \ and\ \bibinfo {author} {\bibfnamefont {G.-B.}\
  \bibnamefont {Jo}},\ }\href
  {http://advances.sciencemag.org/content/4/2/eaao4748} {\bibfield  {journal}
  {\bibinfo  {journal} {Science Advances}\ }\textbf {\bibinfo {volume} {4}},\
  \bibinfo {pages} {2, eaao4748} (\bibinfo {year} {2018})}\BibitemShut
  {NoStop}%
\bibitem [{\citenamefont {de~L\'es\'eleuc}\ \emph {et~al.}(2018)\citenamefont
  {de~L\'es\'eleuc}, \citenamefont {Lienhard}, \citenamefont {Scholl},
  \citenamefont {Barredo}, \citenamefont {Weber}, \citenamefont {Lang},
  \citenamefont {B\"uchler}, \citenamefont {Lahaye},\ and\ \citenamefont
  {Browaeys}}]{fSPTarxiv2018}%
  \BibitemOpen
  \bibfield  {author} {\bibinfo {author} {\bibfnamefont {S.}~\bibnamefont
  {de~L\'es\'eleuc}}, \bibinfo {author} {\bibfnamefont {V.}~\bibnamefont
  {Lienhard}}, \bibinfo {author} {\bibfnamefont {P.}~\bibnamefont {Scholl}},
  \bibinfo {author} {\bibfnamefont {D.}~\bibnamefont {Barredo}}, \bibinfo
  {author} {\bibfnamefont {S.}~\bibnamefont {Weber}}, \bibinfo {author}
  {\bibfnamefont {N.}~\bibnamefont {Lang}}, \bibinfo {author} {\bibfnamefont
  {H.~P.}\ \bibnamefont {B\"uchler}}, \bibinfo {author} {\bibfnamefont
  {T.}~\bibnamefont {Lahaye}}, \ and\ \bibinfo {author} {\bibfnamefont
  {A.}~\bibnamefont {Browaeys}},\ }\href@noop {} {} (\bibinfo {year} {2018}),\
  \Eprint {http://arxiv.org/abs/arXiv:1810.13286} {arXiv:1810.13286}
  \BibitemShut {NoStop}%
\bibitem [{\citenamefont {Yan}\ \emph {et~al.}(2013)\citenamefont {Yan},
  \citenamefont {Moses}, \citenamefont {Gadway}, \citenamefont {Covey},
  \citenamefont {Hazzard}, \citenamefont {Rey}, \citenamefont {Jin},\ and\
  \citenamefont {Ye}}]{Yan:2013xe}%
  \BibitemOpen
  \bibfield  {author} {\bibinfo {author} {\bibfnamefont {B.}~\bibnamefont
  {Yan}}, \bibinfo {author} {\bibfnamefont {S.~A.}\ \bibnamefont {Moses}},
  \bibinfo {author} {\bibfnamefont {B.}~\bibnamefont {Gadway}}, \bibinfo
  {author} {\bibfnamefont {J.~P.}\ \bibnamefont {Covey}}, \bibinfo {author}
  {\bibfnamefont {K.~R.~A.}\ \bibnamefont {Hazzard}}, \bibinfo {author}
  {\bibfnamefont {A.~M.}\ \bibnamefont {Rey}}, \bibinfo {author} {\bibfnamefont
  {D.~S.}\ \bibnamefont {Jin}}, \ and\ \bibinfo {author} {\bibfnamefont
  {J.}~\bibnamefont {Ye}},\ }\href@noop {} {\bibfield  {journal} {\bibinfo
  {journal} {Nature}\ }\textbf {\bibinfo {volume} {501}},\ \bibinfo {pages}
  {521} (\bibinfo {year} {2013})}\BibitemShut {NoStop}%
\bibitem [{\citenamefont {Hazzard}\ \emph {et~al.}(2014)\citenamefont
  {Hazzard}, \citenamefont {Gadway}, \citenamefont {Foss-Feig}, \citenamefont
  {Yan}, \citenamefont {Moses}, \citenamefont {Covey}, \citenamefont {Yao},
  \citenamefont {Lukin}, \citenamefont {Ye}, \citenamefont {Jin},\ and\
  \citenamefont {Rey}}]{PhysRevLett.113.195302}%
  \BibitemOpen
  \bibfield  {author} {\bibinfo {author} {\bibfnamefont {K.~R.~A.}\
  \bibnamefont {Hazzard}}, \bibinfo {author} {\bibfnamefont {B.}~\bibnamefont
  {Gadway}}, \bibinfo {author} {\bibfnamefont {M.}~\bibnamefont {Foss-Feig}},
  \bibinfo {author} {\bibfnamefont {B.}~\bibnamefont {Yan}}, \bibinfo {author}
  {\bibfnamefont {S.~A.}\ \bibnamefont {Moses}}, \bibinfo {author}
  {\bibfnamefont {J.~P.}\ \bibnamefont {Covey}}, \bibinfo {author}
  {\bibfnamefont {N.~Y.}\ \bibnamefont {Yao}}, \bibinfo {author} {\bibfnamefont
  {M.~D.}\ \bibnamefont {Lukin}}, \bibinfo {author} {\bibfnamefont
  {J.}~\bibnamefont {Ye}}, \bibinfo {author} {\bibfnamefont {D.~S.}\
  \bibnamefont {Jin}}, \ and\ \bibinfo {author} {\bibfnamefont {A.~M.}\
  \bibnamefont {Rey}},\ }\href {\doibase 10.1103/PhysRevLett.113.195302}
  {\bibfield  {journal} {\bibinfo  {journal} {Phys. Rev. Lett.}\ }\textbf
  {\bibinfo {volume} {113}},\ \bibinfo {pages} {195302} (\bibinfo {year}
  {2014})}\BibitemShut {NoStop}%
\bibitem [{\citenamefont {Gorshkov}\ \emph {et~al.}(2011)\citenamefont
  {Gorshkov}, \citenamefont {Manmana}, \citenamefont {Chen}, \citenamefont
  {Ye}, \citenamefont {Demler}, \citenamefont {Lukin},\ and\ \citenamefont
  {Rey}}]{PhysRevLett.107.115301}%
  \BibitemOpen
  \bibfield  {author} {\bibinfo {author} {\bibfnamefont {A.~V.}\ \bibnamefont
  {Gorshkov}}, \bibinfo {author} {\bibfnamefont {S.~R.}\ \bibnamefont
  {Manmana}}, \bibinfo {author} {\bibfnamefont {G.}~\bibnamefont {Chen}},
  \bibinfo {author} {\bibfnamefont {J.}~\bibnamefont {Ye}}, \bibinfo {author}
  {\bibfnamefont {E.}~\bibnamefont {Demler}}, \bibinfo {author} {\bibfnamefont
  {M.~D.}\ \bibnamefont {Lukin}}, \ and\ \bibinfo {author} {\bibfnamefont
  {A.~M.}\ \bibnamefont {Rey}},\ }\href {\doibase
  10.1103/PhysRevLett.107.115301} {\bibfield  {journal} {\bibinfo  {journal}
  {Phys. Rev. Lett.}\ }\textbf {\bibinfo {volume} {107}},\ \bibinfo {pages}
  {115301} (\bibinfo {year} {2011})}\BibitemShut {NoStop}%
\bibitem [{\citenamefont {de~Paz}\ \emph {et~al.}(2013)\citenamefont {de~Paz},
  \citenamefont {Sharma}, \citenamefont {Chotia}, \citenamefont {Mar\'echal},
  \citenamefont {Huckans}, \citenamefont {Pedri}, \citenamefont {Santos},
  \citenamefont {Gorceix}, \citenamefont {Vernac},\ and\ \citenamefont
  {Laburthe-Tolra}}]{PhysRevLett.111.185305}%
  \BibitemOpen
  \bibfield  {author} {\bibinfo {author} {\bibfnamefont {A.}~\bibnamefont
  {de~Paz}}, \bibinfo {author} {\bibfnamefont {A.}~\bibnamefont {Sharma}},
  \bibinfo {author} {\bibfnamefont {A.}~\bibnamefont {Chotia}}, \bibinfo
  {author} {\bibfnamefont {E.}~\bibnamefont {Mar\'echal}}, \bibinfo {author}
  {\bibfnamefont {J.~H.}\ \bibnamefont {Huckans}}, \bibinfo {author}
  {\bibfnamefont {P.}~\bibnamefont {Pedri}}, \bibinfo {author} {\bibfnamefont
  {L.}~\bibnamefont {Santos}}, \bibinfo {author} {\bibfnamefont
  {O.}~\bibnamefont {Gorceix}}, \bibinfo {author} {\bibfnamefont
  {L.}~\bibnamefont {Vernac}}, \ and\ \bibinfo {author} {\bibfnamefont
  {B.}~\bibnamefont {Laburthe-Tolra}},\ }\href {\doibase
  10.1103/PhysRevLett.111.185305} {\bibfield  {journal} {\bibinfo  {journal}
  {Phys. Rev. Lett.}\ }\textbf {\bibinfo {volume} {111}},\ \bibinfo {pages}
  {185305} (\bibinfo {year} {2013})}\BibitemShut {NoStop}%
\bibitem [{\citenamefont {Yao}\ \emph {et~al.}(2018)\citenamefont {Yao},
  \citenamefont {Zaletel}, \citenamefont {Stamper-Kurn},\ and\ \citenamefont
  {Vishwanath}}]{DSL2015}%
  \BibitemOpen
  \bibfield  {author} {\bibinfo {author} {\bibfnamefont {N.~Y.}\ \bibnamefont
  {Yao}}, \bibinfo {author} {\bibfnamefont {M.~P.}\ \bibnamefont {Zaletel}},
  \bibinfo {author} {\bibfnamefont {D.~M.}\ \bibnamefont {Stamper-Kurn}}, \
  and\ \bibinfo {author} {\bibfnamefont {A.}~\bibnamefont {Vishwanath}},\
  }\href {\doibase 10.1038/s41567-017-0030-7} {\bibfield  {journal} {\bibinfo
  {journal} {Nature Physics}\ }\textbf {\bibinfo {volume} {14}},\ \bibinfo
  {pages} {405} (\bibinfo {year} {2018})}\BibitemShut {NoStop}%
\bibitem [{\citenamefont {Zou}\ \emph {et~al.}(2017)\citenamefont {Zou},
  \citenamefont {Zhao},\ and\ \citenamefont {Liu}}]{Our2017}%
  \BibitemOpen
  \bibfield  {author} {\bibinfo {author} {\bibfnamefont {H.}~\bibnamefont
  {Zou}}, \bibinfo {author} {\bibfnamefont {E.}~\bibnamefont {Zhao}}, \ and\
  \bibinfo {author} {\bibfnamefont {W.~V.}\ \bibnamefont {Liu}},\ }\href
  {\doibase 10.1103/PhysRevLett.119.050401} {\bibfield  {journal} {\bibinfo
  {journal} {Phys. Rev. Lett.}\ }\textbf {\bibinfo {volume} {119}},\ \bibinfo
  {pages} {050401} (\bibinfo {year} {2017})}\BibitemShut {NoStop}%
\bibitem [{\citenamefont {Kele\ifmmode~\mbox{\c{s}}\else \c{s}\fi{}}\ and\
  \citenamefont {Zhao}(2018{\natexlab{a}})}]{keles2018absence}%
  \BibitemOpen
  \bibfield  {author} {\bibinfo {author} {\bibfnamefont {A.}~\bibnamefont
  {Kele\ifmmode~\mbox{\c{s}}\else \c{s}\fi{}}}\ and\ \bibinfo {author}
  {\bibfnamefont {E.}~\bibnamefont {Zhao}},\ }\href {\doibase
  10.1103/PhysRevLett.120.187202} {\bibfield  {journal} {\bibinfo  {journal}
  {Phys. Rev. Lett.}\ }\textbf {\bibinfo {volume} {120}},\ \bibinfo {pages}
  {187202} (\bibinfo {year} {2018}{\natexlab{a}})}\BibitemShut {NoStop}%
\bibitem [{\citenamefont {Kele\ifmmode~\mbox{\c{s}}\else \c{s}\fi{}}\ and\
  \citenamefont {Zhao}(2018{\natexlab{b}})}]{keles-prb}%
  \BibitemOpen
  \bibfield  {author} {\bibinfo {author} {\bibfnamefont {A.}~\bibnamefont
  {Kele\ifmmode~\mbox{\c{s}}\else \c{s}\fi{}}}\ and\ \bibinfo {author}
  {\bibfnamefont {E.}~\bibnamefont {Zhao}},\ }\href {\doibase
  10.1103/PhysRevB.97.245105} {\bibfield  {journal} {\bibinfo  {journal} {Phys.
  Rev. B}\ }\textbf {\bibinfo {volume} {97}},\ \bibinfo {pages} {245105}
  (\bibinfo {year} {2018}{\natexlab{b}})}\BibitemShut {NoStop}%
\bibitem [{\citenamefont {Zou}\ \emph {et~al.}(2019)\citenamefont {Zou},
  \citenamefont {Zhao}, \citenamefont {Guan},\ and\ \citenamefont
  {Liu}}]{Our2019}%
  \BibitemOpen
  \bibfield  {author} {\bibinfo {author} {\bibfnamefont {H.}~\bibnamefont
  {Zou}}, \bibinfo {author} {\bibfnamefont {E.}~\bibnamefont {Zhao}}, \bibinfo
  {author} {\bibfnamefont {X.-W.}\ \bibnamefont {Guan}}, \ and\ \bibinfo
  {author} {\bibfnamefont {W.~V.}\ \bibnamefont {Liu}},\ }\href {\doibase
  10.1103/PhysRevLett.122.180401} {\bibfield  {journal} {\bibinfo  {journal}
  {Phys. Rev. Lett.}\ }\textbf {\bibinfo {volume} {122}},\ \bibinfo {pages}
  {180401} (\bibinfo {year} {2019})}\BibitemShut {NoStop}%
\bibitem [{\citenamefont {Ueda}\ and\ \citenamefont
  {Onoda}(2014{\natexlab{b}})}]{Ueda2014Chiral}%
  \BibitemOpen
  \bibfield  {author} {\bibinfo {author} {\bibfnamefont {H.}~\bibnamefont
  {Ueda}}\ and\ \bibinfo {author} {\bibfnamefont {S.}~\bibnamefont {Onoda}},\
  }\href {\doibase 10.1103/PhysRevB.89.024407} {\bibfield  {journal} {\bibinfo
  {journal} {Phys. Rev. B}\ }\textbf {\bibinfo {volume} {89}},\ \bibinfo
  {pages} {024407} (\bibinfo {year} {2014}{\natexlab{b}})}\BibitemShut
  {NoStop}%
\bibitem [{\citenamefont {Luttinger}\ and\ \citenamefont
  {Tisza}(1946)}]{LuttingerTisza}%
  \BibitemOpen
  \bibfield  {author} {\bibinfo {author} {\bibfnamefont {J.~M.}\ \bibnamefont
  {Luttinger}}\ and\ \bibinfo {author} {\bibfnamefont {L.}~\bibnamefont
  {Tisza}},\ }\href {\doibase 10.1103/PhysRev.70.954} {\bibfield  {journal}
  {\bibinfo  {journal} {Phys. Rev.}\ }\textbf {\bibinfo {volume} {70}},\
  \bibinfo {pages} {954} (\bibinfo {year} {1946})}\BibitemShut {NoStop}%
\bibitem [{\citenamefont {Pollmann}\ \emph {et~al.}(2012)\citenamefont
  {Pollmann}, \citenamefont {Berg}, \citenamefont {Turner},\ and\ \citenamefont
  {Oshikawa}}]{Pollmann2012}%
  \BibitemOpen
  \bibfield  {author} {\bibinfo {author} {\bibfnamefont {F.}~\bibnamefont
  {Pollmann}}, \bibinfo {author} {\bibfnamefont {E.}~\bibnamefont {Berg}},
  \bibinfo {author} {\bibfnamefont {A.~M.}\ \bibnamefont {Turner}}, \ and\
  \bibinfo {author} {\bibfnamefont {M.}~\bibnamefont {Oshikawa}},\ }\href
  {\doibase 10.1103/PhysRevB.85.075125} {\bibfield  {journal} {\bibinfo
  {journal} {Phys. Rev. B}\ }\textbf {\bibinfo {volume} {85}},\ \bibinfo
  {pages} {075125} (\bibinfo {year} {2012})}\BibitemShut {NoStop}%
\bibitem [{\citenamefont {den Nijs}\ and\ \citenamefont
  {Rommelse}(1989)}]{string1989}%
  \BibitemOpen
  \bibfield  {author} {\bibinfo {author} {\bibfnamefont {M.}~\bibnamefont {den
  Nijs}}\ and\ \bibinfo {author} {\bibfnamefont {K.}~\bibnamefont {Rommelse}},\
  }\href@noop {} {\bibfield  {journal} {\bibinfo  {journal} {Phys. Rev. B}\
  }\textbf {\bibinfo {volume} {40}},\ \bibinfo {pages} {4709} (\bibinfo {year}
  {1989})}\BibitemShut {NoStop}%
\bibitem [{\citenamefont {Majumdar}\ and\ \citenamefont
  {Ghosh}(1969)}]{Majumdar1969}%
  \BibitemOpen
  \bibfield  {author} {\bibinfo {author} {\bibfnamefont {C.~K.}\ \bibnamefont
  {Majumdar}}\ and\ \bibinfo {author} {\bibfnamefont {D.~K.}\ \bibnamefont
  {Ghosh}},\ }\href {\doibase 10.1063/1.1664979} {\bibfield  {journal}
  {\bibinfo  {journal} {Journal of Mathematical Physics}\ }\textbf {\bibinfo
  {volume} {10}},\ \bibinfo {pages} {1399} (\bibinfo {year}
  {1969})}\BibitemShut {NoStop}%
\bibitem [{\citenamefont {Shastry}\ and\ \citenamefont
  {Sutherland}(1981)}]{Shastry1981}%
  \BibitemOpen
  \bibfield  {author} {\bibinfo {author} {\bibfnamefont {B.~S.}\ \bibnamefont
  {Shastry}}\ and\ \bibinfo {author} {\bibfnamefont {B.}~\bibnamefont
  {Sutherland}},\ }\href {\doibase 10.1103/PhysRevLett.47.964} {\bibfield
  {journal} {\bibinfo  {journal} {Phys. Rev. Lett.}\ }\textbf {\bibinfo
  {volume} {47}},\ \bibinfo {pages} {964} (\bibinfo {year} {1981})}\BibitemShut
  {NoStop}%
\bibitem [{\citenamefont {Kanter}(1989)}]{Kanter}%
  \BibitemOpen
  \bibfield  {author} {\bibinfo {author} {\bibfnamefont {I.}~\bibnamefont
  {Kanter}},\ }\href {\doibase 10.1103/PhysRevB.39.7270} {\bibfield  {journal}
  {\bibinfo  {journal} {Phys. Rev. B}\ }\textbf {\bibinfo {volume} {39}},\
  \bibinfo {pages} {7270} (\bibinfo {year} {1989})}\BibitemShut {NoStop}%
\bibitem [{\citenamefont {Giamarchi}(2003)}]{Giamarchi2003}%
  \BibitemOpen
  \bibfield  {author} {\bibinfo {author} {\bibfnamefont {T.}~\bibnamefont
  {Giamarchi}},\ }\href {\doibase 10.1093/acprof:oso/9780198525004.001.0001}
  {\emph {\bibinfo {title} {Quantum Physics in One Dimension}}}\ (\bibinfo
  {publisher} {Oxford University Press},\ \bibinfo {year} {2003})\BibitemShut
  {NoStop}%
\end{thebibliography}
%
\end{document}